\documentclass[numberedappendix]{emulateapj}

\shorttitle{Faint Variables in IC\,1613}
\shortauthors{Bernard et al.}

\begin{document}

\title{The ACS LCID Project. II. Faint Variable Stars \\
       in the Isolated Dwarf Irregular Galaxy IC\,1613\altaffilmark{1}}

\author{Edouard J. Bernard,\altaffilmark{2,3,4}
    Matteo Monelli,\altaffilmark{2,3}
    Carme Gallart,\altaffilmark{2,3}
    Antonio Aparicio,\altaffilmark{2,3}
    Santi Cassisi,\altaffilmark{5} \\
    Igor Drozdovsky,\altaffilmark{2,3,6}
    Sebastian L. Hidalgo,\altaffilmark{2,3}
    Evan D. Skillman,\altaffilmark{7}
    Peter B. Stetson\altaffilmark{8}}

\altaffiltext{1}{Based on observations made with the NASA/ESA {\it Hubble Space
    Telescope}, obtained at the Space Telescope Science Institute, which is
    operated by the Association of Universities for Research in Astronomy,
    Inc., under NASA contract NAS5-26555. These observations are associated
    with program 10505.}
\altaffiltext{2}{Instituto de Astrof\'{i}sica de Canarias, La Laguna, Tenerife,
    Spain; monelli@iac.es, carme@iac.es, antapaj@iac.es, dio@iac.es,
    slhidalgo@iac.es.}
\altaffiltext{3}{Departamento de Astrof\'{i}sica, Universidad de La Laguna,
    Tenerife, Spain.}
\altaffiltext{4}{Current address: Institute for Astronomy, Royal Observatory,
    University of Edinburgh, UK; ejb@roe.ac.uk.}
\altaffiltext{5}{INAF-Osservatorio Astronomico di Collurania,
    Teramo, Italy; cassisi@oa-teramo.inaf.it.}
\altaffiltext{6}{Astronomical Institute, St. Petersburg State University,
    St. Petersburg, Russia.}
\altaffiltext{7}{Department of Astronomy, University of Minnesota,
    Minneapolis, USA; skillman@astro.umn.edu.}
\altaffiltext{8}{Dominion Astrophysical Observatory, Herzberg Institute of
    Astrophysics, National Research Council, Victoria, Canada;
    peter.stetson@nrc-cnrc.gc.ca.}

\begin{abstract}

 We present the results of a new search for variable stars in the Local Group
 (LG) isolated dwarf galaxy IC\,1613, based on 24 orbits of F475W and F814W
 photometry from the {\it Advanced Camera for Surveys} onboard the {\it Hubble
 Space Telescope}. We detected 259 candidate variables in this field, of which
 only 13 (all of them bright Cepheids) were previously known. Out of the
 confirmed variables, we found 90 RR~Lyrae stars, 49 classical Cepheids
 (including 36 new discoveries), and 38 eclipsing binary stars for which we
 could determine a period.
 The RR~Lyrae include 61 fundamental (RR$ab$) and 24 first-overtone (RR$c$)
 pulsators, and 5 pulsating in both modes simultaneously (RR$d$). As for the
 majority of LG dwarfs, the mean periods of the RR$ab$ and RR$c$ (0.611 and
 0.334 day, respectively) as well as the fraction of overtone pulsators
 (f$_c$=0.28) place this galaxy in the intermediate regime between the
 Oosterhoff types.
 From their position on the period-luminosity diagram and light-curve
 morphology, we can unambiguously classify 25 and 14 Cepheids as fundamental
 and first-overtone mode pulsators, respectively. Another two are clearly
 second-overtone Cepheids, the first ones to be discovered beyond the
 Magellanic Clouds.
 Among the remaining candidate variables, five were classified as
 $\delta$-Scuti and five as long-period variables. Most of the others are
 located on the main-sequence, the majority of them likely eclipsing binary
 systems, although some present variations similar to pulsating stars.
 We estimate the distance to IC\,1613 using various methods based on the
 photometric and pulsational properties of the Cepheids and RR~Lyrae stars. The
 values we find are in very good agreement with each other and with previous
 estimates based on independent methods. When corrected to a common reddening
 of E(B$-$V)=0.025 and true LMC distance modulus of
 (m$-$M)$_{LMC,0}$=18.515$\pm$0.085, we find that all the distance
 determinations from the literature converge to a common value of
 (m$-$M)$_0$=24.400$\pm$0.014 (statistical), or 760 kpc.
 The parallel WFPC2 field, which lies within three core radii, was also searched
 for variable stars. We discovered nine RR Lyrae stars (4 RR$ab$, 4 RR$c$ and
 1 RR$d$) and two Cepheids, even though the lower signal-to-noise ratio of the
 observations did not allow us to measure their periods as accurately as for the
 variables in the ACS field-of-view. We provide their coordinates and
 approximate properties for completeness.

\end{abstract}


\section{Introduction}

 The search for variable stars in IC\,1613 started at the beginning of the
 1930's with the seminal work of Baade, Hubble, and Mayall, although it was
 published only later in \citet{san71}. The discovery of 37 definite Cepheids
 allowed Baade to calculate the first distance to IC\,1613 based on the
 period-luminosity (PL) relation. The analysis of the faint Cepheids in the
 same photographic plates was completed by \citet{car90} and the PL relation
 extended to periods as short as about two days.
 Candidate RR~Lyrae stars were also observed in IC\,1613 from ground-
 \citep{sah92} and space-based imaging \citep{dol01}, confirming the
 presence of a bona fide old population.
 More recent surveys by \citet[and references therein]{man01} and the OGLE
 collaboration \citep{uda01} of a central $\sim$14$\arcmin$x14$\arcmin$ field
 led to the discovery of 128 and 138 Cepheids, respectively, although the small
 size of the telescopes used by both groups limited the sample to relatively
 bright, long-period Cepheids. While all the Cepheids with periods longer than
 about 10~days in IC\,1613 have most likely been discovered, very few have been
 found at the faint end of the PL relation \citep[e.g.,][]{dol01}.

 Here we present the analysis of very deep {\it Hubble Space Telescope}
 ({\it HST}) ACS data with the aim of searching for short-period variable
 stars. These data are part of a larger set obtained in the context of the LCID
 project\footnotemark[9] (C. Gallart et al. 2010, in preparation).
\footnotetext[9]{Local Cosmology from Isolated Dwarfs:
    http://www.iac.es/project/LCID/.}
 The depth and quality of the data also allow the reconstruction of the full
 star formation history (SFH) of this field located just outside of the core
 radius. The study of the stellar populations and SFH from the color-magnitude
 diagram (CMD) fitting technique will be presented in a companion paper
 (E. Skillman et al. 2010, in preparation).

 The structure of the present paper follows closely that of
 \citet[hereafter Paper~I]{ber09}
 in which we analysed the properties of the variables in the dSph galaxies Cetus
 and Tucana. A summary of the observations and data reduction is presented in
 \S~\ref{sec:2}, while \S~\ref{sec:3} and \S~\ref{sec:4} deal with the
 identification of variable stars and the sample completeness, respectively. In
 \S~\ref{sec:5} we describe the sample of RR~Lyrae stars and the Cepheids are
 presented in \S~\ref{sec:6}. The eclipsing binaries and other candidate
 variables are presented in \S~\ref{sec:7} to \ref{sec:9}. Our list of variables
 is cross-identified with the previous catalogs covering this field in
 \S~\ref{sec:10}. In \S~\ref{sec:11} we use the properties of the RR~Lyrae stars
 and Cepheids to estimate the distance to IC\,1613 and compare it with the
 values of the literature. In the last section we discuss the results and
 present our conclusions.

    \defcitealias{ber09}{Paper~I}

\begin{figure*}
\epsscale{1.1} 
\plotone{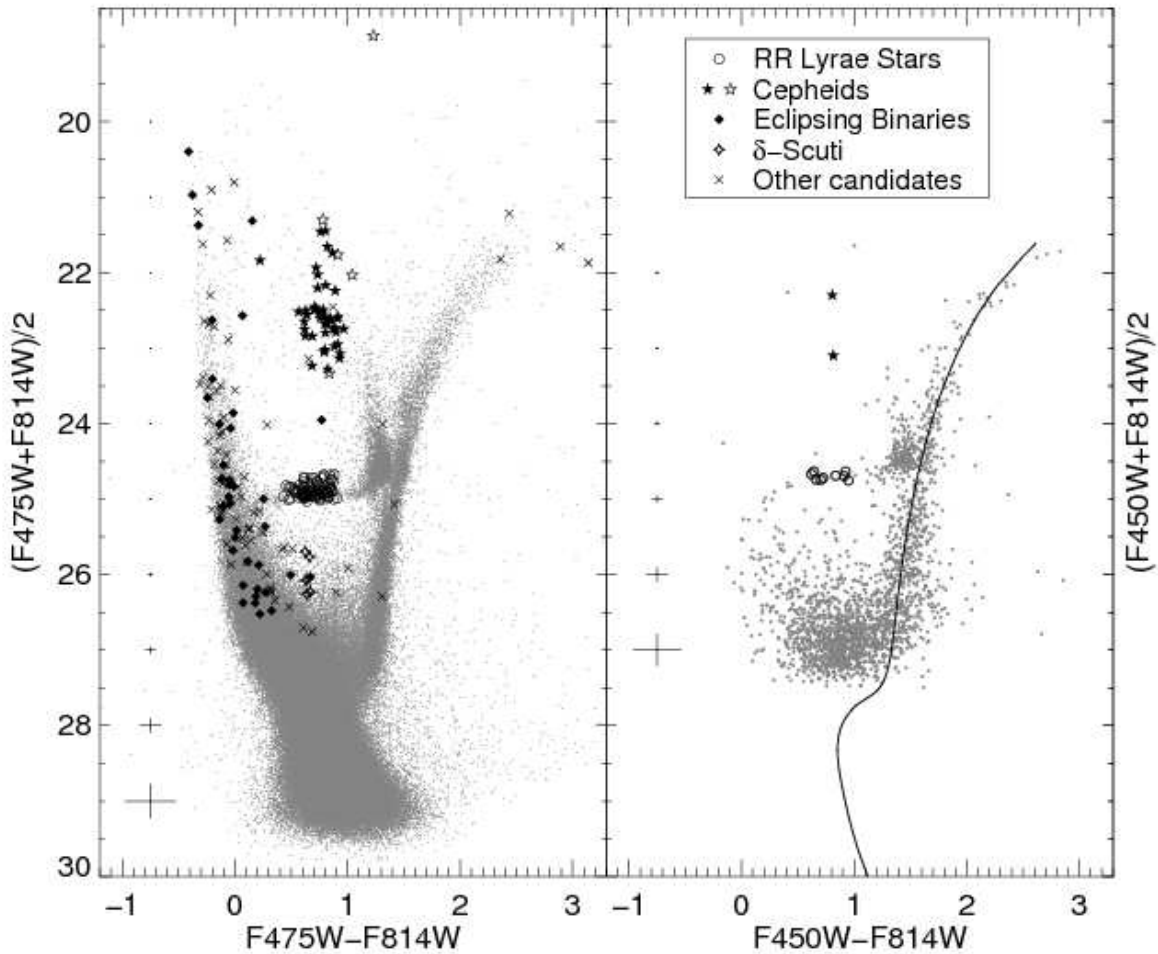}
\figcaption{Color-magnitude diagrams of IC\,1613 for the ACS ({\it left}) and
 WFPC2 ({\it right}) fields, where the confirmed and candidate variables have
 been overplotted, as labeled in the inset.
 An isochrone from the BaSTI library \citep[Z=0.0003, 13\,Gyr,][]{pie04} is
 overlaid in the right panel to show the location of the old main-sequence
 turn-off.
\label{fig:1}}
\end{figure*}


\begin{deluxetable}{ccccc}
\tabletypesize{\scriptsize}
\tablewidth{0pt}
\tablecaption{Observing Log\label{tab1}}
\tablehead{
\colhead{Date} & \colhead{UT Start} & \colhead{MHJD\tablenotemark{a}} &
\colhead{Filter} & \colhead{Exp. Time}}
\startdata
 2006 Aug 18  &  06:59:48  &  53965.304727  &  F475W  &  1325 \\
 2006 Aug 18  &  07:24:48  &  53965.320827  &  F814W  &  1106 \\
 2006 Aug 18  &  08:35:20  &  53965.370932  &  F475W  &  1300 \\
 2006 Aug 18  &  08:59:55  &  53965.387147  &  F814W  &  1153 \\
 2006 Aug 18  &  10:11:37  &  53965.437935  &  F475W  &  1324 \\
 2006 Aug 18  &  10:36:36  &  53965.454023  &  F814W  &  1107 \\
 2006 Aug 18  &  11:47:09  &  53965.504139  &  F475W  &  1300 \\
 2006 Aug 18  &  12:11:44  &  53965.520354  &  F814W  &  1153 \\
 2006 Aug 18  &  13:23:26  &  53965.571142  &  F475W  &  1324 \\
 2006 Aug 18  &  13:48:25  &  53965.587230  &  F814W  &  1107 \\
 2006 Aug 18  &  14:58:58  &  53965.637346  &  F475W  &  1300 \\
 2006 Aug 18  &  15:23:33  &  53965.653565  &  F814W  &  1153 \\
 2006 Aug 18  &  16:35:15  &  53965.704349  &  F475W  &  1324 \\
 2006 Aug 18  &  17:00:14  &  53965.720437  &  F814W  &  1107 \\
 2006 Aug 18  &  18:10:47  &  53965.770553  &  F475W  &  1300 \\
 2006 Aug 18  &  18:35:22  &  53965.786769  &  F814W  &  1153 \\
 2006 Aug 19  &  06:58:25  &  53966.303776  &  F475W  &  1324 \\
 2006 Aug 19  &  07:23:24  &  53966.319864  &  F814W  &  1107 \\
 2006 Aug 19  &  08:33:57  &  53966.369980  &  F475W  &  1300 \\
 2006 Aug 19  &  08:58:32  &  53966.386197  &  F814W  &  1153 \\
 2006 Aug 19  &  10:10:14  &  53966.436983  &  F475W  &  1324 \\
 2006 Aug 19  &  10:35:13  &  53966.453071  &  F814W  &  1107 \\
 2006 Aug 19  &  11:45:46  &  53966.503187  &  F475W  &  1300 \\
 2006 Aug 19  &  12:10:21  &  53966.519403  &  F814W  &  1153 \\
 2006 Aug 19  &  13:22:02  &  53966.570179  &  F475W  &  1324 \\
 2006 Aug 19  &  13:47:02  &  53966.586278  &  F814W  &  1107 \\
 2006 Aug 19  &  14:57:34  &  53966.636383  &  F475W  &  1300 \\
 2006 Aug 19  &  15:22:10  &  53966.652610  &  F814W  &  1153 \\
 2006 Aug 19  &  16:33:52  &  53966.703397  &  F475W  &  1324 \\
 2006 Aug 19  &  16:58:51  &  53966.719486  &  F814W  &  1107 \\
 2006 Aug 19  &  18:09:24  &  53966.769602  &  F475W  &  1300 \\
 2006 Aug 19  &  18:33:59  &  53966.785817  &  F814W  &  1153 \\
 2006 Aug 20  &  05:21:08  &  53967.236226  &  F475W  &  1324 \\
 2006 Aug 20  &  05:46:07  &  53967.252314  &  F814W  &  1107 \\
 2006 Aug 20  &  06:56:41  &  53967.302442  &  F475W  &  1300 \\
 2006 Aug 20  &  07:21:16  &  53967.318658  &  F814W  &  1153 \\
 2006 Aug 20  &  08:32:57  &  53967.369434  &  F475W  &  1324 \\
 2006 Aug 20  &  08:57:56  &  53967.385520  &  F814W  &  1107 \\
 2006 Aug 20  &  10:08:30  &  53967.435649  &  F475W  &  1300 \\
 2006 Aug 20  &  10:33:05  &  53967.451865  &  F814W  &  1153 \\
 2006 Aug 20  &  11:44:46  &  53967.502641  &  F475W  &  1324 \\
 2006 Aug 20  &  12:09:45  &  53967.518729  &  F814W  &  1107 \\
 2006 Aug 20  &  13:20:19  &  53967.568857  &  F475W  &  1300 \\
 2006 Aug 20  &  13:44:54  &  53967.585072  &  F814W  &  1153 \\
 2006 Aug 20  &  14:56:35  &  53967.635848  &  F475W  &  1324 \\
 2006 Aug 20  &  15:21:34  &  53967.651936  &  F814W  &  1107 \\
 2006 Aug 20  &  16:32:08  &  53967.702064  &  F475W  &  1300 \\
 2006 Aug 20  &  16:56:43  &  53967.718279  &  F814W  &  1153
\enddata
\tablenotetext{a}{Modified Heliocentric Julian Date of mid-exposure: HJD$ - $2,400,000.}
\end{deluxetable}


\section{Observations and Data Reduction}\label{sec:2}

\subsection{Primary ACS Imaging}

 This work is based on observations obtained with the ACS onboard the {\it HST}
 of a field located about 5$\arcmin$ West of the center of IC\,1613. The field
 location was chosen to be far enough out in radius to minimize crowding effects
 for high quality photometry, yet close enough in to maximize the total number
 of stars observed.
 As the goal of these observations was to reach the oldest main sequence
 turn-offs with good photometric accuracy, 24 {\it HST} orbits were devoted to
 this galaxy. These were collected over about 2.4 consecutive days between 2006
 August 28 and 30.
 Each orbit was split in two $\sim$1200 seconds exposures in F475W and F814W
 for an optimal sampling of the light curves. The complete observing log is
 presented in Table~\ref{tab1}.

 The DAOPHOT/ALLFRAME suite of programs \citep{ste94} was used to obtain the
 instrumental photometry of the stars on the individual, non-drizzled images
 provided by the {\it HST} pipeline.
 Additionally, we used the pixel area maps and data quality masks to correct for
 the variations of the pixel areas on the sky around the field and to flag bad
 pixels.
 Standard calibration was carried out as described in \citet{sir05}, taking
 into account the updated zero-points of \citet{mac07} following the lowering
 of the Wide Field Channel temperature setpoint in July 2006.
 We refer the reader to \citet{mon10} for a detailed description of the
 data reduction and calibration. The final CMD is presented in Fig.~\ref{fig:1}
 ({\it left panel}), where the $(F475W+F814W)/2 \sim V$ filter combination was
 chosen for the ordinate axis so that the horizontal-branch (HB) appears
 approximately horizontal.

 The light-curves of the Cepheids and RR~Lyrae stars were converted to Johnson
 BVI magnitudes using the transformations given in \citetalias{ber09}. Since 12
 of our Cepheids were also observed by the OGLE team in V and I (see
 \S~\ref{sec:10}), we could check the accuracy of our transformations. Ten
 out of the 12 stars in common have well sampled light-curves and accurately
 determined periods, although 6 of these only have V observations in the OGLE
 database because of their relatively low luminosities.

\begin{figure}
\epsscale{1.0} 
\plotone{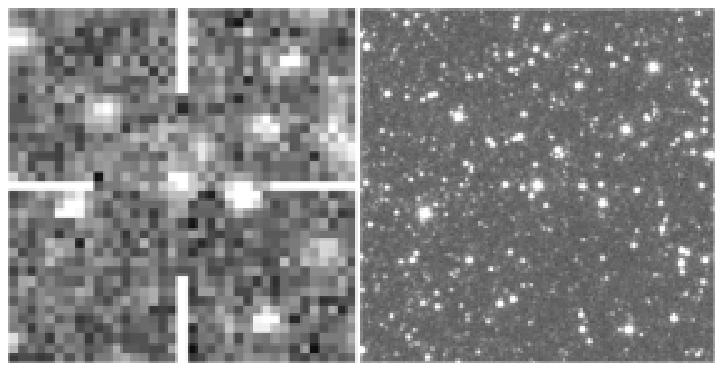}
\plotone{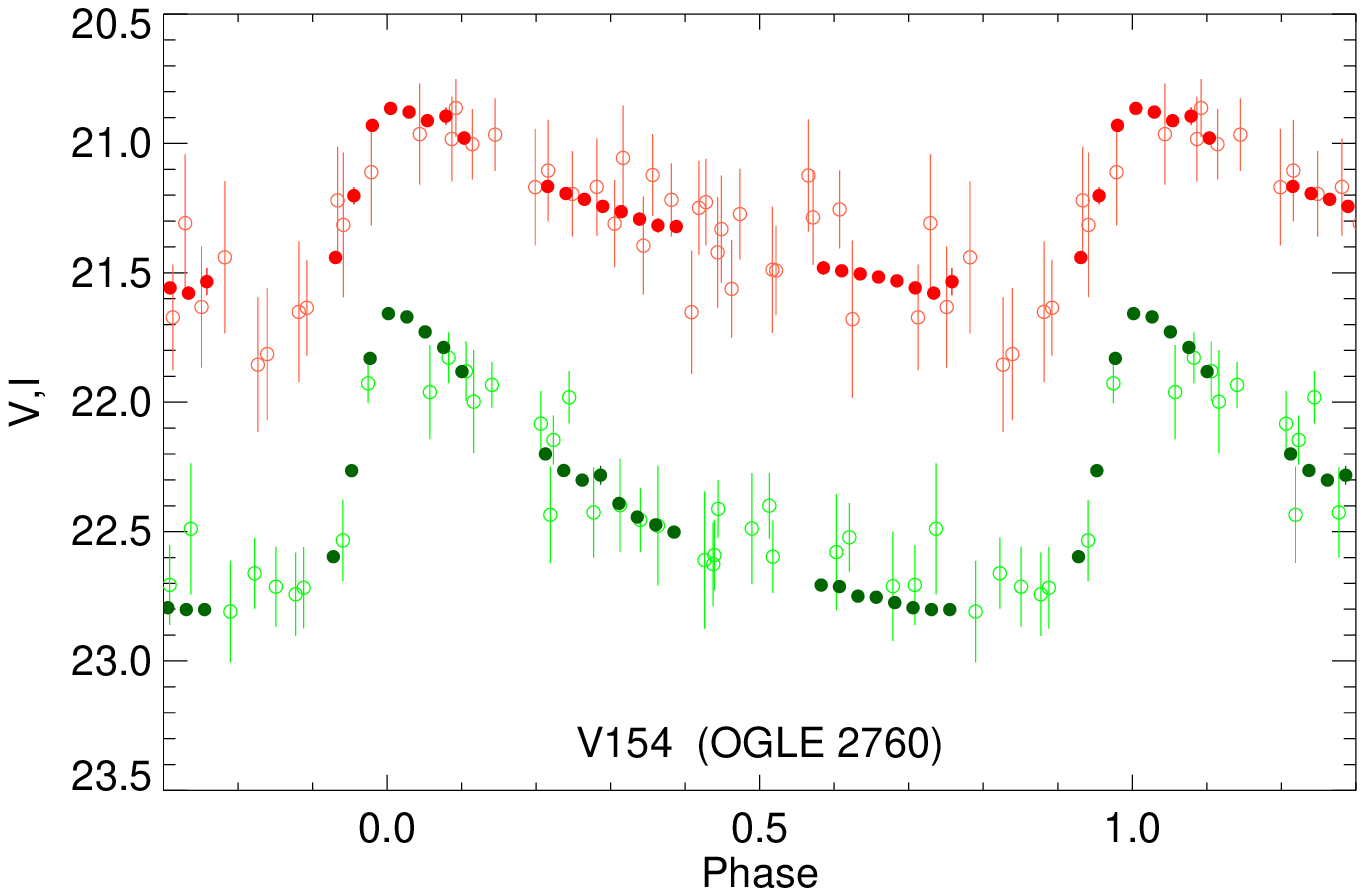}
\plotone{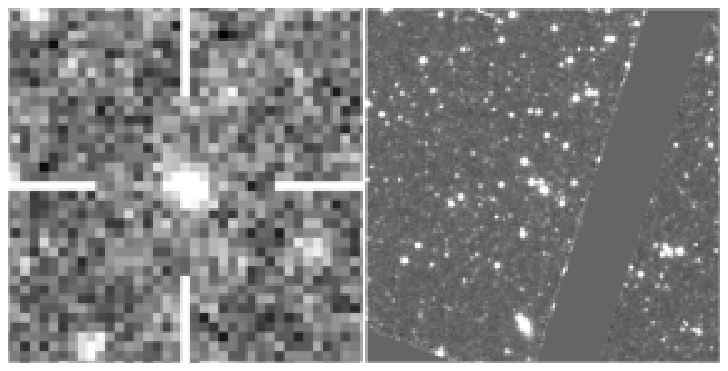}
\plotone{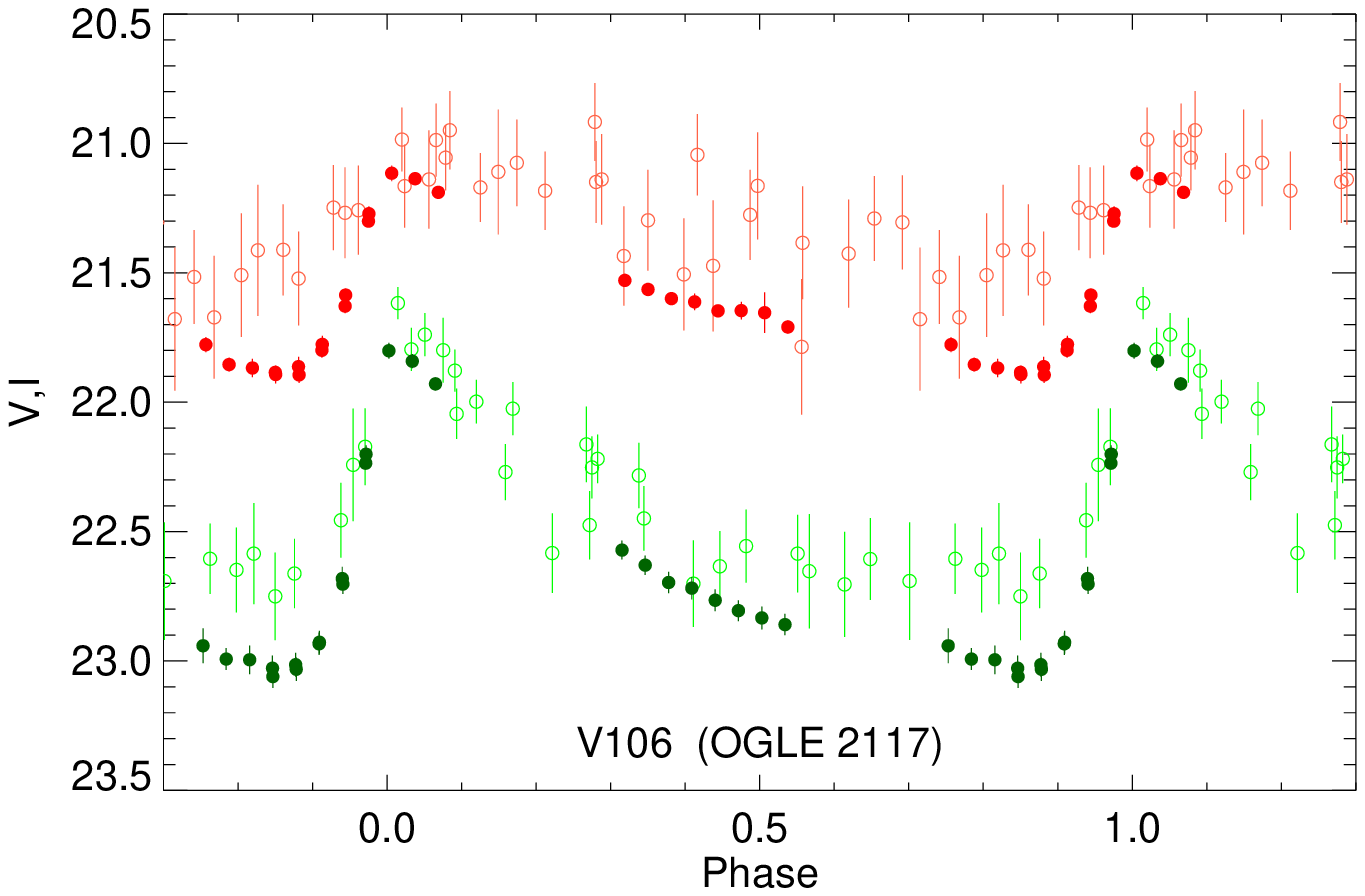}
\figcaption{Comparison of the finding charts and Johnson photometry from the
 OGLE database with ours for V154 ({\it top}) and V106 ({\it bottom}).
 The finding charts are 15$\arcsec$ on a side. The OGLE light curves are
 represented as open symbols, while our photometry is shown as filled symbols.
 For clarity, the V-band datapoints have been shifted downward by 0.2 mag.
 Note the good agreement of the ground-based and {\it HST} light-curve in the
 first case. The second is an example of ground-based photometry affected by
 unresolved bright companions, which affects the light-curve morphology and
 mean-magnitude.
\label{fig:2}}
\end{figure}

 We find excellent agreement at all phases between the two photometry sets for 7
 of the 10 usable stars. The remaining three present a shift in magnitude which
 ranges from $\sim$0.1 at maximum light to $\sim$0.5 for the faintest points.
 The comparison of the finding charts shows that this difference is due to
 contamination by nearby bright stars that are not resolved in the ground-based
 data. To illustrate this, in Fig.~\ref{fig:2} we compare the
 finding-charts and photometry of the OGLE database with ours for two Cepheids
 (V106 and V154).
 For each Cepheid, we show 15$\arcsec$x15$\arcsec$ stamps from the OGLE I-band
 and our F814W images, as well as the V and I light-curves from both
 photometries.
 While both Cepheids are located in rather crowded fields, the stars close to
 V154 are relatively faint compared to the Cepheid. The ground-based and
 {\it HST} light-curves are therefore overlapping perfectly. On the other hand,
 in the case of V106 the neighbours have a luminosity similar to that of the
 Cepheid, thus affecting the mean magnitude and amplitude of the OGLE
 light-curve.

 This test shows that our photometry in the Johnson bands, although transformed
 from a very different photometric system, is very reliable and can be safely
 used to analyse the properties of the variable stars, both in terms of
 magnitude and amplitude. This is emphasized in \S~\ref{sec:11} where the
 distance to IC\,1613 is calculated from Johnson luminosities and for which we
 find excellent agreement with previous measurements based on other methods.


\begin{deluxetable}{ccc|ccc}
\tablewidth{0pt}
\tablecaption{Photometry of the Variable and Candidate Variable Stars in IC\,1613 -- HST Bands\label{tab2}}
\tablehead{
\colhead{MHJD\tablenotemark{a}} & \colhead{$m_{475}$} & \colhead{$\sigma_{475}$} &
\colhead{MHJD\tablenotemark{a}} & \colhead{$m_{814}$} & \colhead{$\sigma_{814}$}}
\startdata
\multicolumn{6}{c}{V001} \\ \hline
 53965.304727 & 24.978 & 0.036 &  53965.320827 & 24.420 & 0.026 \\
 53965.370932 & 25.216 & 0.043 &  53965.387147 & 24.410 & 0.039 \\
 53965.437935 & 25.356 & 0.035 &  53965.454023 & 24.583 & 0.030 \\
 53965.504139 & 25.500 & 0.038 &  53965.520354 & 24.634 & 0.042 \\
 53965.571142 & 25.562 & 0.033 &  53965.587230 & 24.669 & 0.052
\enddata
\tablecomments{Table \ref{tab2} is published in its entirety in the
electronic edition of {\it The Astrophysical Journal}.  A portion is
shown here for guidance regarding its form and content.}
\tablenotetext{a}{Modified Heliocentric Julian Date of mid-exposure: HJD$ - $2,400,000.}
\end{deluxetable}


\subsection{Parallel WFPC2 Imaging}


\begin{deluxetable*}{ccc|ccc|ccc} 

\tablewidth{0pt}
\tablecaption{Photometry of the Variable and Candidate Variable Stars in IC\,1613 -- Johnson Bands\label{tab3}}
\tablehead{
\colhead{MHJD\tablenotemark{a}} & \colhead{$B$} & \colhead{$\sigma_B$} &
\colhead{MHJD\tablenotemark{a}} & \colhead{$V$} & \colhead{$\sigma_V$} &
\colhead{MHJD\tablenotemark{a}} & \colhead{$I$} & \colhead{$\sigma_I$}}
\startdata
\multicolumn{9}{c}{V001} \\ \hline
 53965.304727 & 25.054 & 0.036 &  53965.312777 & 24.789 & 0.036 &  53965.320827 & 24.403 & 0.026 \\
 53965.370932 & 25.337 & 0.043 &  53965.379040 & 24.939 & 0.043 &  53965.387147 & 24.392 & 0.039 \\
 53965.437935 & 25.471 & 0.035 &  53965.445979 & 25.090 & 0.035 &  53965.454023 & 24.565 & 0.030 \\
 53965.504139 & 25.633 & 0.038 &  53965.512247 & 25.202 & 0.038 &  53965.520354 & 24.618 & 0.042 \\
 53965.571142 & 25.701 & 0.033 &  53965.579186 & 25.255 & 0.033 &  53965.587230 & 24.654 & 0.052
\enddata
\tablecomments{Table \ref{tab3} is published in its entirety in the
electronic edition of {\it The Astrophysical Journal}.  A portion is
shown here for guidance regarding its form and content.}
\tablenotetext{a}{Modified Heliocentric Julian Date of mid-exposure: HJD$ - $2,400,000.}
\end{deluxetable*}               

\begin{deluxetable*}{cccccccccccccccccc} 

\tablewidth{0pt}
\tablecaption{Properties of Variable Stars in IC\,1613 -- ACS Field\label{tab4}}
\tablehead{
\colhead{} & \colhead{} & \colhead{R.A.} & \colhead{Decl.} & \colhead{Period} &
\colhead{} & \colhead{} & \colhead{} & \colhead{} & \colhead{} &
\colhead{$\langle F475W \rangle -$}\\
\colhead{ID} & \colhead{Type} & \colhead{(J2000)} & \colhead{(J2000)} &
\colhead{(days)} & \colhead{log P} &
\colhead{$\langle F475W \rangle$} & \colhead{$A_{475}$} &
\colhead{$\langle F814W \rangle$} & \colhead{$A_{814}$} &
\colhead{$\langle F814W \rangle$} &
\colhead{$\langle B \rangle$} & \colhead{$A_B$} &
\colhead{$\langle V \rangle$} & \colhead{$A_V$} &
\colhead{$\langle I \rangle$} & \colhead{$A_I$}}
\startdata
  V001 & $ab$ &  01 04 20.90 &  02 10 36.8 &      0.592  &      $-$0.228  &      25.272  &  1.170      &      24.562  &      0.555  &         0.710  &      25.374  &      1.280  &      25.037  &      0.890  &      24.545  &      0.558  \\
  V002 &  $c$ &  01 04 20.96 &  02 10 27.4 &      0.425  &      $-$0.372  &      25.076  &  0.465      &      24.459  &      0.239  &         0.618  &      25.163  &      0.508  &      24.866  &      0.381  &      24.442  &      0.237  \\
  V003 &  EB  &  01 04 20.99 &  02 09 41.6 &      2.900  &     ~~\,0.462  & {\it 24.341} & {\it 0.168} & {\it 23.572} & {\it 0.107} & {\it    0.769} & {\it 24.449} & {\it 0.254} & {\it 24.073} & {\it 0.169} & {\it 23.549} & {\it 0.146} \\
  V004 &  $c$ &  01 04 22.03 &  02 09 10.7 &      0.314  &      $-$0.503  &      25.051  &  0.361      &      24.595  &      0.167  &         0.455  &      25.110  &      0.390  &      24.898  &      0.287  &      24.583  &      0.163  \\
  V005 & $ab$ &  01 04 22.10 &  02 09 13.1 &      0.484  &      $-$0.315  &      25.237  &  0.978      &      24.616  &      0.544  &         0.621  &      25.347  &      1.111  &      24.995  &      0.809  &      24.599  &      0.541
\enddata
\tablecomments{Table \ref{tab4} is published in its entirety in the
electronic edition of {\it The Astrophysical Journal}.  A portion is
shown here for guidance regarding its form and content.}

\end{deluxetable*}


 IC\,1613 was also observed with the WFPC2 in the F450W and F814W bands as
 parallel exposures to the primary ACS observations. This provided the
 opportunity to analyze a second field with the same number of observations and
 a similar exposure time. The orientation of the ACS field was chosen such that
 the parallel WFPC2 field would sample a region further from the center of the
 galaxy ($\sim$11$\arcmin$) in order to study population gradients.
 The images from the Wide Field chips were reduced individually as described in
 \citet{tur97}, and calibrated following the instructions from the {\it HST}
 Data Handbook for WFPC2.\footnotemark[10] Given the small number of stars on
 the Planetary Camera, the photometry of this chip was performed on the stacked
 images only.
\footnotetext[10]{http://www.stsci.edu/instruments/wfpc2/Wfpc2\_dhb/wfpc2\_ ch52.html}
 The resulting CMD for the whole WFPC2 field-of-view is shown in the right panel
 of Fig.~\ref{fig:1}, where a 13~Gyr old isochrone from the BaSTI library
 \citep{pie04} has been overplotted assuming $A_B$=0.123 and $A_I$=0.056
 \citep{sch98} for the F450W and F814W bands, respectively, and a dereddened
 distance modulus of 24.40 (see \S~\ref{sec:11.3}).

 Unfortunately, due to the lower sensitivity of the instrument with respect to
 the ACS, especially in the F450W band, the CMD does not reach the oldest
 main-sequence turn-offs. However, the lack of bright main-sequence stars in
 this outer field reveals the presence of a blue horizontal-branch component,
 typical of old, low metallicity stellar populations.

\section{Identification of Variable Stars}\label{sec:3}

 The candidate variables were extracted from the photometric catalogs using the
 variability index of \citet{wel93}. This process yielded $\sim$780 candidates
 in the primary field.
 A preliminary check of the light-curve and position on the CMD, together with
 a careful inspection of the stacked image, allowed us to discard false
 detections due to cosmic-ray hits, chip defects or stars located under the
 wings of bright stars. This left us with 259 candidates; these are shown in
 Fig.~\ref{fig:1} overplotted on the CMD using their intensity-averaged
 magnitudes when available (see below). The individual F475W and F814W
 measurements for all of the candidate variables are listed in
 Table~\ref{tab2}, while the transformed BVI magnitudes are given in
 Table~\ref{tab3}.

 The period search was performed on the suspected variables through Fourier
 analysis \citep{hor86} taking into account the information from both bands
 simultaneously, then refined interactively by modifying the period until a
 tighter light curve was obtained.
 For each variable, datapoints with error bars larger than 3-$\sigma$ above the
 mean error bar size were rejected through sigma clipping with five iterations.
 As the period-finding program is interactive, it was possible to selectively
 reject more or less datapoints depending on the light curve quality before
 recalculating the periodogram. However, except in few particular cases, we
 found that the period-search was not affected by a few bad points.
 Given the short timebase of the observations
 ($<$3~days), the periods are given with three significant figures only.
 The accuracy of period determination was estimated from the degradation of the
 quality of the light curves when applying small period offsets. It mainly
 depends on the period itself and on the time interval covered by observations,
 and ranges from about 0.001 day for the shorter period RR~Lyrae stars to few
 hundredths of a day for the longest period Cepheids.

 The classification of the candidates was based on their light-curve morphology
 and position in the CMD.
 We found 90 RR~Lyrae stars, 49 Cepheids, 38 eclipsing binaries, and five
 candidate $\delta$-Scuti stars.
 A significant number of variable candidates was also found along the main
 sequence. For most of these, however, we could not find a period that would
 produce a smooth light-curve. In the following, we will simply refer to these
 variables as main-sequences variables (MSV), although most of them are
 probably eclipsing binaries.

 To obtain the amplitudes and intensity-averaged magnitudes of the monoperiodic
 variables in the instability strip, we fitted the light-curves with a set of
 templates based on the set of \citet[][see \citetalias{ber09}]{lay99}.
 The amplitudes of the double-mode RR Lyrae stars were measured from a low-order
 Fourier fit to the light-curve phased with the primary period after
 prewhitening of the secondary period.
 The mean magnitude and color of the variables for which we could not find a
 period are weighted averages from the ALLFRAME photometry and are therefore
 only approximate. Table~\ref{tab4} summarizes the properties of the confirmed
 variable stars in the ACS field. The first two columns give the identification
 number and type of variability, while the next two list the equatorial
 coordinates (J2000.0). Columns (5) and (6) give the primary period in days,
 i.e., the first-overtone period in the case of the RR$d$, and the logarithm of
 this period. The intensity-averaged magnitudes $\langle F475W \rangle$ and
 $\langle F814W \rangle$, and color $\langle F475W \rangle- \langle F814W
 \rangle$ are given in columns (7), (9), and (11). The amplitudes in the F475W
 and F814W bands measured from the template fits are listed in the eighth and
 tenth columns. The last six columns alternately list the intensity-averaged
 magnitudes and amplitudes in the Johnson B, V, and I bands. Approximate values
 are listed in italics.
 A list of the remaining candidates, providing the coordinates and approximate
 magnitudes, is given in Table~\ref{tab5}. They are labeled as MSV, long-period
 variables (LPV), classical instability strip variables (ISV), and RGB variables
 (RGV) based on their location on the CMD.

 The same procedure was followed with the WFPC2 photometry, in which we found
 11 variables: nine RR~Lyrae stars and two Cepheids. However, the low
 signal-to-noise of the individual measurements produced rather noisy
 light-curves, and some low amplitude variables might have been missed.
 Given the small number of variables in the parallel field and the generally
 lower quality of their photometry and inferred parameters, we only give
 their coordinates and approximate parameters in Table~\ref{tab6} for
 completeness. These variables will not be taken into account when calculating
 average properties of the IC\,1613 variable star population.


\begin{deluxetable}{lcccccccccc}  

\tablewidth{0pt}
\tablecaption{Candidate Variable Stars in IC\,1613 -- ACS Field\label{tab5}}
\tablehead{
\colhead{} & \colhead{} & \colhead{R.A.} & \colhead{Decl.} &
\colhead{} & \colhead{}\\
\colhead{ID} & \colhead{Type} & \colhead{(J2000)} & \colhead{(J2000)} &
\colhead{$F475W$} & \colhead{$F814W$}}
\startdata
  VC194 &  MSV &  01 04 20.27 &  02 10 28.6 &  25.272 &  25.261  \\
  VC195 &  RGV &  01 04 20.62 &  02 10 01.1 &  25.777 &  24.361  \\
  VC196 &  MSV &  01 04 22.63 &  02 09 53.7 &  26.401 &  26.078  \\
  VC197 &  MSV &  01 04 23.08 &  02 08 06.6 &  26.680 &  25.784  \\
  VC198 &  MSV &  01 04 23.24 &  02 08 33.6 &  24.092 &  24.221
\enddata
\tablecomments{Table \ref{tab5} is published in its entirety in the
electronic edition of {\it The Astrophysical Journal}.  A portion is
shown here for guidance regarding its form and content.}
\end{deluxetable}

\begin{deluxetable*}{lcccccccccc}  

\tablewidth{0pt}
\tablecaption{Properties of Variable Stars in IC\,1613 -- WFPC2 Field\label{tab6}}
\tablehead{
\colhead{} & \colhead{} & \colhead{R.A.} & \colhead{Decl.} & \colhead{Period} &
\colhead{} & \colhead{} & \colhead{} & \colhead{$\langle F450W \rangle -$}\\
\colhead{ID} & \colhead{Type} & \colhead{(J2000)} & \colhead{(J2000)} &
\colhead{(days)} & \colhead{log P} &
\colhead{$\langle F450W \rangle$} & \colhead{$\langle F814W \rangle$} &
\colhead{$\langle F814W \rangle$} &
\colhead{$A_{450}$} & \colhead{$A_{814}$}}
\startdata
 V183  &   $d$  &  01 04 04.90  &  02 13 44.9  &  0.34  &  $-$0.46  &  25.178  &  24.446  &  0.732  &  0.463  &  0.146 \\
 V184  &   $c$  &  01 04 05.05  &  02 13 11.0  &  0.36  &  $-$0.44  &  24.955  &  24.313  &  0.642  &  0.520  &  0.299 \\
 V185  &  $ab$  &  01 04 05.20  &  02 13 32.7  &  0.56  &  $-$0.25  &  25.150  &  24.240  &  0.910  &  1.168  &  0.551 \\
 V186  &   $c$  &  01 04 05.29  &  02 11 01.7  &  0.35  &  $-$0.46  &  25.107  &  24.272  &  0.834  &  0.510  &  0.234 \\
 V187  &  $ab$  &  01 04 07.61  &  02 13 00.8  &  0.55  &  $-$0.26  &  25.096  &  24.393  &  0.703  &  1.615  &  0.565 \\
 V188  &  $ab$  &  01 04 07.99  &  02 11 32.5  &  0.65  &  $-$0.19  &  25.097  &  24.174  &  0.924  &  0.794  &  0.409 \\
 V189  &  $ab$  &  01 04 08.65  &  02 11 11.0  &  0.60  &  $-$0.22  &  25.228  &  24.279  &  0.949  &  0.773  &  0.296 \\
 V190  &   Cep  &  01 04 10.25  &  02 13 06.9  &  0.67  &  $-$0.17  &  23.507  &  22.693  &  0.813  &  0.664  &  0.401 \\
 V191  &   $c$  &  01 04 10.32  &  02 12 50.9  &  0.33  &  $-$0.48  &  24.979  &  24.359  &  0.619  &  0.624  &  0.344 \\
 V192  &   Cep  &  01 04 10.40  &  02 12 03.7  &  1.30  & ~~\,0.11  &  22.701  &  21.896  &  0.805  &  0.657  &  0.285 \\
 V193  &   $c$  &  01 04 10.97  &  02 12 08.1  &  0.36  &  $-$0.44  &  25.072  &  24.411  &  0.661  &  0.516  &  0.276
\enddata

\end{deluxetable*}


\section{Completeness and Areal Coverage}\label{sec:4}

 In this section, we analyze the effects of stellar crowding and
 signal-to-noise (SNR) limitations, temporal sampling, and spatial coverage on
 the completeness of our Cepheids and RR~Lyrae stars samples.

 The high spatial resolution of the ACS and the depth
 of our data imply that incompleteness will become noticeable well below the
 HB. Artificial-star tests (E. Skillman et al. 2010, in preparation)
 indicate that the completeness is higher than 98\% at (F475W+F814W)/2 $\sim$
 25.0. Therefore, down to these magnitudes only variables with amplitude of the
 order of the error bars at this magnitude ($\sim$0.1) might have been missed.
 However, variables fainter than the HB (e.g., SX Phoenicis and/or
 $\delta$-Scuti) have probably been missed due to crowding and low SNR.
 In addition, even though these variables are present in this galaxy (see
 \S~\ref{sec:8}), the relatively long exposure time smoothing out the
 variations in luminosity and the rather slow temporal sampling hampered the
 detection of these short-period variables.

\begin{figure}
\epsscale{1.1} 
\plotone{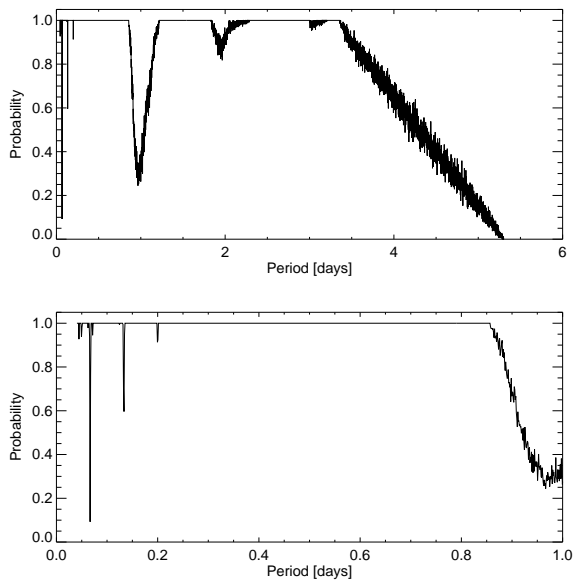}
\figcaption{Probability of detecting variable stars in IC\,1613 as a function
 of period, for periods between about 1~hr and 5~days ({\it top}), and close-up
 view for periods between about 1~hr and 1 day ({\it bottom}).
\label{fig:3}}
\end{figure}

 We also estimate the completeness due to temporal sampling by carrying out
 numerical simulations in the same way as described in \citetalias{ber09}. The
 detection probability is presented in Fig.~\ref{fig:3}.
 In the upper panel, the maximum period that is displayed is limited by the
 observational timebase ($\sim$3 days). Note that even though it is possible to
 detect variable stars with periods longer than the observation time-base, it
 is not feasible to determine their periods.
 The Figure also shows that the probability to detect a variable star in the
 period range of RR~Lyrae stars ($\sim$ 0.2--0.8 day) is basically one. On the
 other hand, a significant dip is present at P$\sim$1 day, due to the
 observations being gathered at the same time of the day each day. While it
 might not prevent detecting variables at this period (most likely Cepheids) if
 their amplitude is larger than about $\sim$0.1, it seriously limits our
 capacity to find an accurate period (e.g., V150 in Fig.~\ref{fig:8}, V030 in
 \S~\ref{sec:6.4}).

\begin{figure*}
\epsscale{1.1} 
\plotone{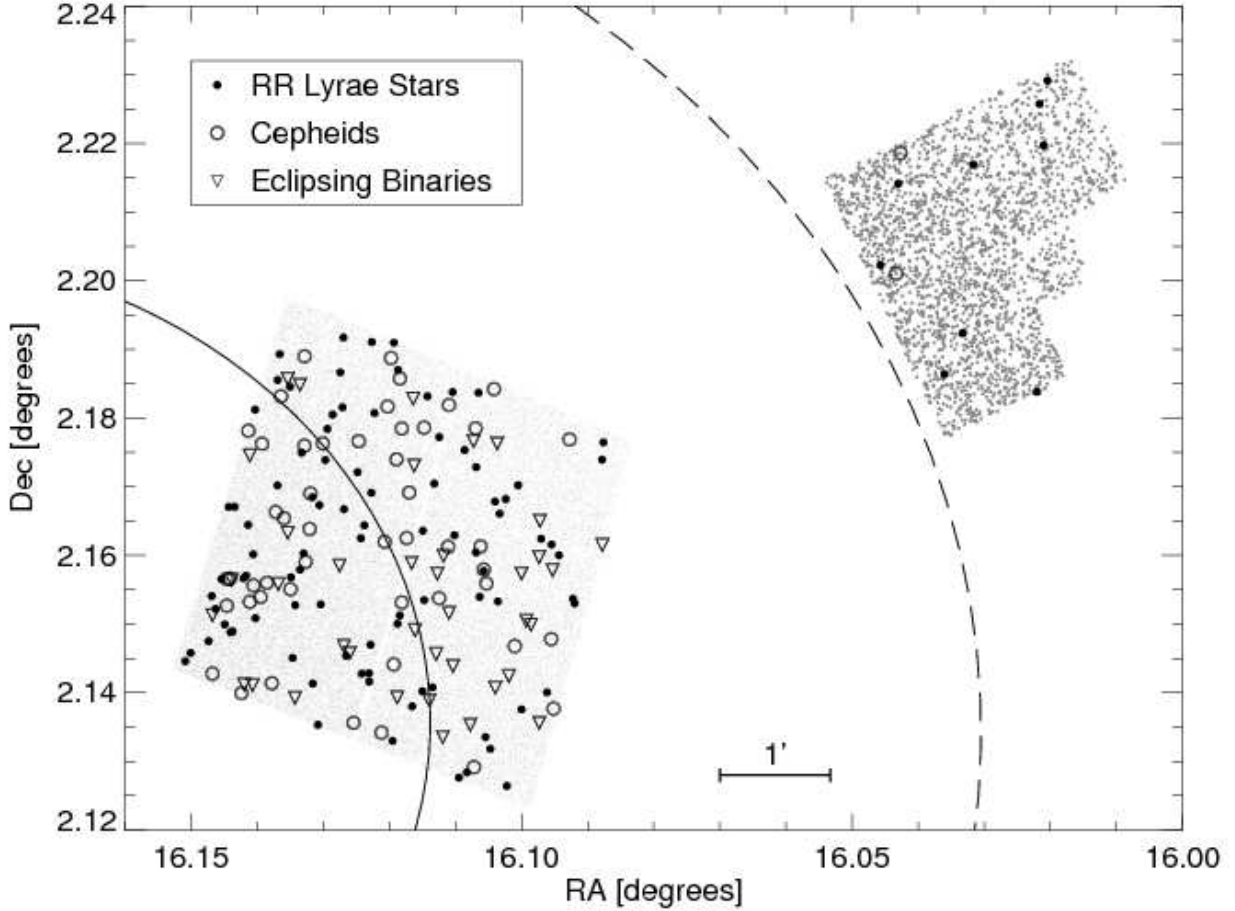}
\figcaption{Spatial distribution of stars in the ACS and WFPC2 fields, where the
 variable stars have been overplotted as labeled in the inset.
 Solid- and dashed-line ellipses represent the core radius
 \citep[$r_c=4.5\arcmin \pm 0.9$,][]{bat07} and twice the core radius.
\label{fig:4}}
\end{figure*}

\begin{figure*}
\epsscale{0.9} 
\plotone{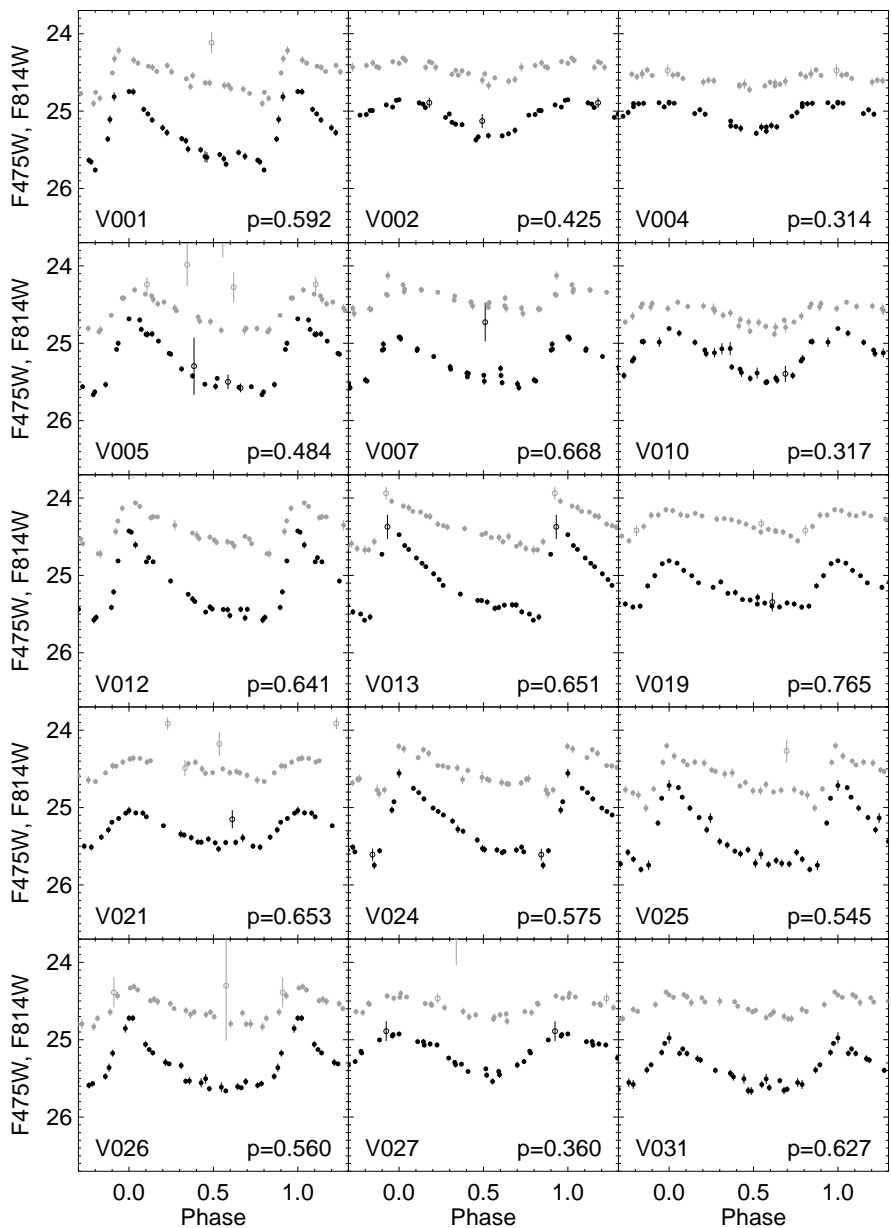}
\figcaption{Light-curves of the RR~Lyrae variables of the ACS field in the
 F475W ({\it black}) and F814W ({\it grey}) bands, phased with the period in
 days shown in the lower right corner of each panel.
 Photometric error bars are shown. The open circles show bad data points, i.e.,
 with errors larger than 3-$\sigma$ above the mean error of a given star, which
 were not used in the calculation of the period and mean magnitudes.
 {\it [Figure~\ref{fig:5} is presented in its entirety in the electronic
 edition of the Astrophysical Journal]}.
\label{fig:5}}
\end{figure*}

\begin{figure}
\epsscale{1.15} 
\plotone{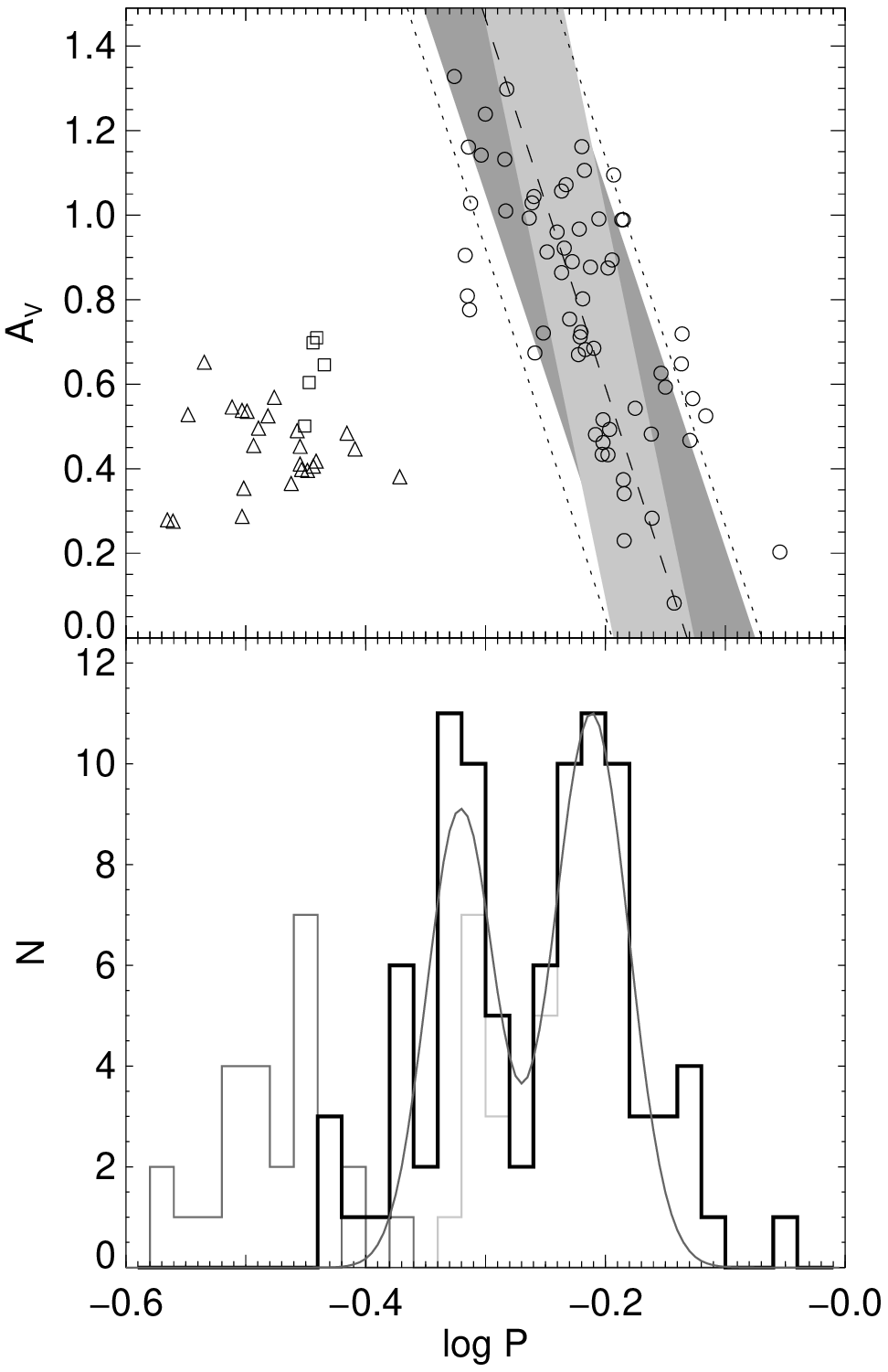}
\figcaption{{\it Top}: Period-amplitude diagram for the RR Lyrae stars.
 Circles, triangles and squares represent RR$ab$, RR$c$, and RR$d$ (plotted
 with their overtone periods) respectively. The dashed line is a fit to the
 period-amplitude of the RR$ab$ after rejecting the points further than
 1.5-$\sigma$ ({\it dotted lines}). The light and dark greyed areas represent
 the $\pm$1.5-$\sigma$ limits of Cetus and Tucana, respectively, from
 \citetalias{ber09}.
 {\it Bottom}: Period histogram for the RR Lyrae stars of the ACS field. RR$ab$
 and RR$c$ are shown as the light and dark gray histograms, respectively,
 while the thick black histogram represent the {\it fundamentalized} RR Lyrae
 stars (RR$ab$, RR$c$, and RR$d$). The solid gray line is a double-Gaussian fit
 to the {\it fundamentalized} histogram.
\label{fig:6}}
\end{figure}

 As expected from the large extent of the galaxy compared to the field-of-view
 of the ACS, the main factor preventing us from obtaining a complete census of
 RR~Lyrae stars and Cepheids in IC\,1613 is the spatial coverage.
 Figure~\ref{fig:4} shows the distribution of stars from our ACS and WFPC2
 fields on top of ellipses representing the core radius \citep[$r_c=4.5\arcmin
 \pm 0.9$,][]{bat07} and twice the core radius. \citet{bat07} estimated a tidal
 radius of $24.4\arcmin \pm 0.3$, ellipticity $\varepsilon = 1-b/a=0.19 \pm
 0.02$, and position angle of 87$\degr$, in good agreement with the values
 found by \citet[][$PA=80\degr$, $\varepsilon=0.15$]{ber07}.
 Thus, the ACS field covers only about 1/160th of the area within the tidal
 radius. If we assume that the distribution of variable stars follows that of
 the overall population, we can estimate the total number of RR~Lyrae stars and
 Cepheids within the tidal radius as described in \citetalias{ber09}.
 Basically, we adopt the shape and orientation of the isodensity contours from
 \citet{bat07} to divide our sample of variable stars into elliptical annuli.
 The area of these annuli was obtained through Monte-Carlo sampling, from which
 we could calculate the density profile of the variable stars. Using this
 profile with the core and tidal radii of \citet{bat07} as input to
 equation~(21) of \citet{kin62}, we estimate that there should be about 2100
 RR~Lyrae stars and 1400 Cepheids within the tidal radius of IC\,1613.
 Therefore, our ACS sample represents about 4\% of the total number of RR~Lyrae
 stars and Cepheids.

\section{RR Lyrae stars}\label{sec:5}

\subsection{ACS sample}

 In our ACS field we detected 90 RR Lyrae stars, all of them new discoveries
 (see \S~\ref{sec:5.3}). From the periods and light-curve shapes, we identified
 61 of them pulsating in the fundamental mode (RR$ab$), 24 in the first-overtone
 mode (RR$c$), and five in both modes simultaneously (RR$d$). The light curves
 of all the RR~Lyrae stars are shown in Fig.~\ref{fig:5}.

 We find the mean period of the RR$ab$ and RR$c$ to be 0.611 and 0.334 day,
 respectively. These values, as well as the fraction of first-overtone
 $f_c=N_c/(N_{ab}+N_c)=0.28$, place IC\,1613 in the gap between the Oosterhoff
 type~I and type~II groups (Oo-I and Oo-II). IC\,1613 therefore joins the
 majority of dwarf galaxies in the Local Group as an Oosterhoff intermediate
 galaxy, the only real exceptions among the dSphs with well-sampled HB to date
 being Ursa Minor \citep{nem88} and Bo\"otes \citep{dal06}.

\begin{figure}
\epsscale{1.2} 
\plotone{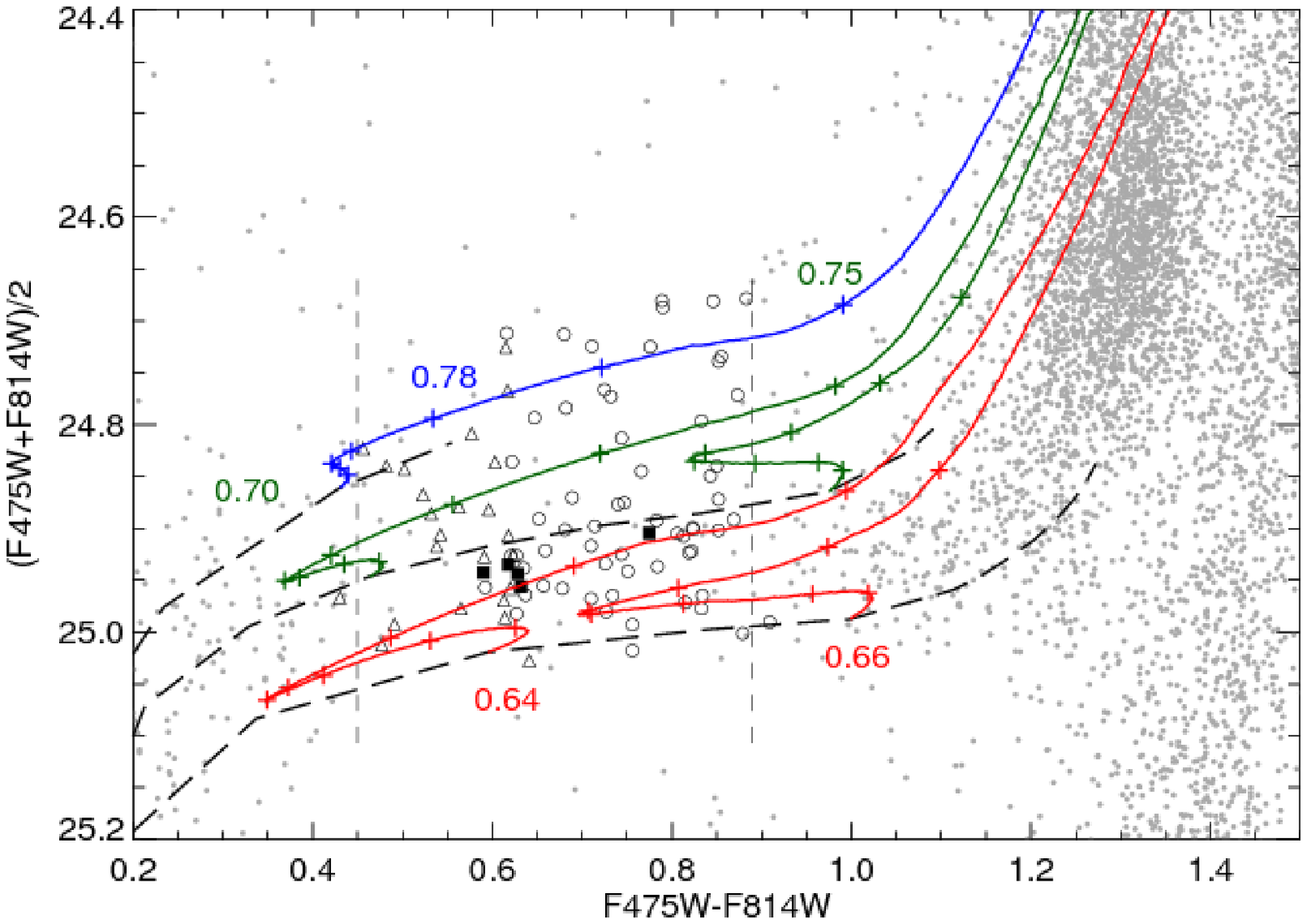}
\figcaption{Zoom-in on the HB of IC\,1613, where the RR Lyrae variables
 have been overplotted. Open circles, open triangles, and filled squares
 represent RR$ab$, RR$c$, and RR$d$, respectively.
 The long-dashed, thick lines represent, from top to bottom, the Z=0.0001,
 0.0003, and 0.001 ZAHB, while the light gray, dark gray, and black lines are
 the corresponding evolutionary tracks with masses as labeled. Overplotted on
 them, the crosses indicate intervals of 10~Myr.
 The vertical dashed lines roughly delimit the instability strip.
\label{fig:7}}
\end{figure}

 In Fig.~\ref{fig:6} we show the period-amplitude diagram ({\it top}) and the
 period distribution ({\it bottom}) of the RR~Lyrae stars. The top panel shows
 that the distribution of the RR$ab$ of IC\,1613 in period-amplitude space is
 very similar to that of Tucana, both in terms of slope and dispersion around
 the fit. The slight shift towards shorter periods of the RR$ab$ stars of
 IC\,1613, compared to those of Tucana, might be explained by the higher
 metallicity of IC\,1613.

\begin{figure*}
\epsscale{0.9} 
\plotone{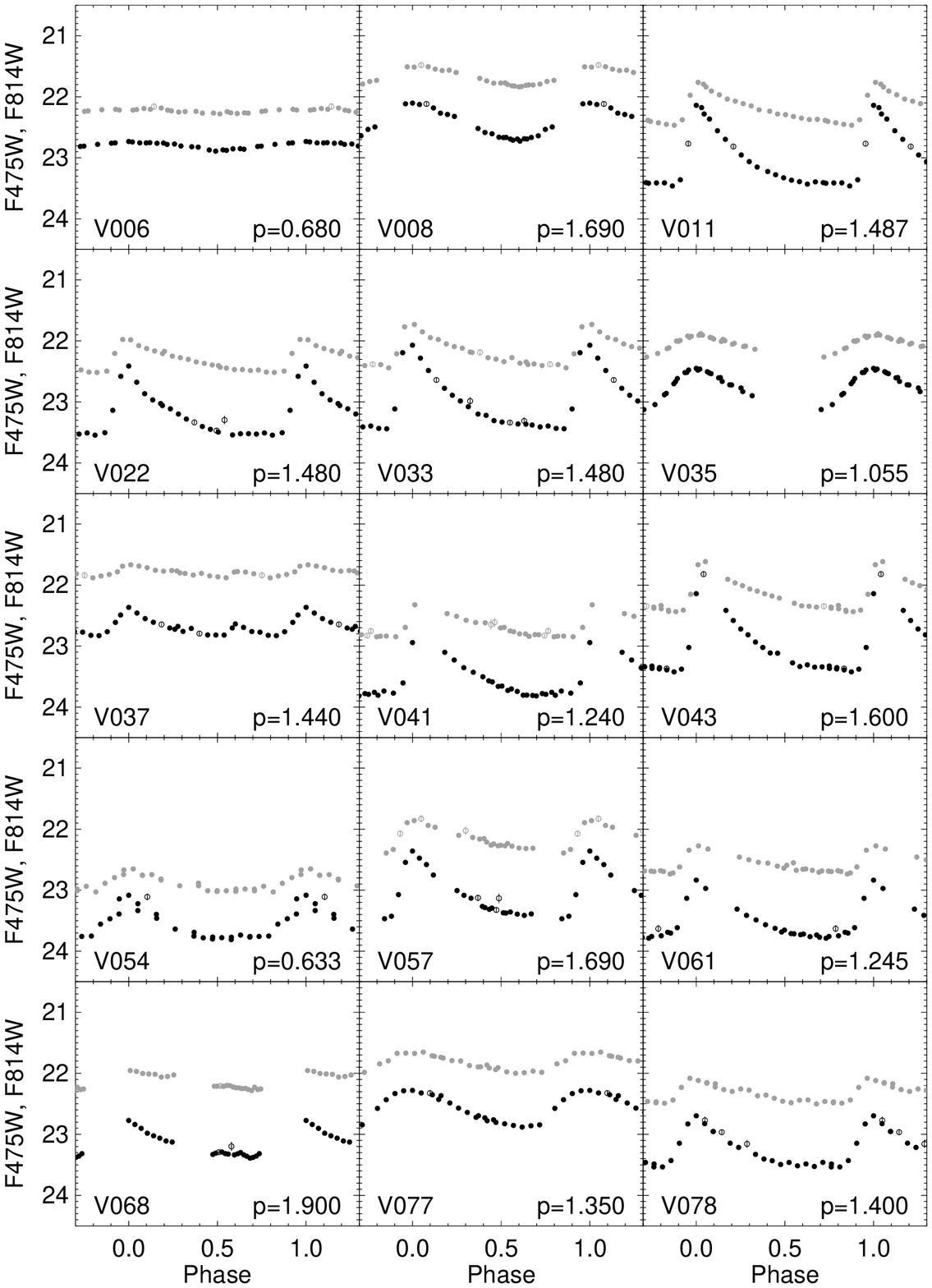}
\figcaption{Light-curves of all the Cepheid variables in the F475W
 ({\it black}) and F814W ({\it grey}) bands.
 Photometric error bars are shown. The open circles
 show the data points with errors larger than 3-$\sigma$ above the mean error
 of a given star. {\it [Figure~\ref{fig:8} is presented in its entirety
 in the electronic edition of the Astrophysical Journal]}.
\label{fig:8}}
\end{figure*}

 However, the distribution of {\it fundamentalized} periods appears to be
 bimodal, as shown by the fit of two Gaussian profiles ({\it solid grey line})
 in the bottom panel, with peaks at P=0.478 and 0.614 day. Given the separation
 between the two peaks and the small uncertainty on the periods
 ($\la$0.001~day), it seems unlikely to be caused by stochastic effects only.
 To check the reliability of the observed bimodality, we applied the KMM test
 of \citet{ash94} to the sample of RR$ab$ periods. The algorithm estimates the
 improvement of a two-group fit over a single Gaussian, and returned a
 $p$-value of 0.075. This can be interpreted as the rejection of the single
 Gaussian model at a confidence level of 92.5\%, and therefore indicates that a
 bimodal distribution is a statistically significant improvement over the
 single Gaussian.

 Note that the distribution of periods of the RR$ab$ alone is also bimodal,
 although the small number of RR$ab$ in the short-period peak does not exclude
 a result due to small number statistics.
 In addition, the location of the peaks (at P=0.494 and 0.615 day) does not seem
 to correspond to the superposition of Oo-I and Oo-II samples, as is the case in
 NGC\,1835 \citep[P$\sim$0.54 and 0.65 day;][]{sos03}.

 The bimodality might nontheless be due to the presence of two distinct old
 populations, as was already observed in the case of Tucana \citep{ber08}.
 Figure~\ref{fig:7} presents a zoom-in of the CMD centered on the HB. Zero-age
 horizontal-branch (ZAHB) loci and tracks from the scaled-solar models of the
 BaSTI library \citep{pie04} are also shown, plotted assuming a reddening of
 E(B$-$V)=0.025 from \citet{sch98}, and a distance modulus of 24.49. This
 distance takes into account the true distance modulus (m$-$M)$_0$=24.44
 determined in \S~\ref{sec:11.3} based on these data and a shift of $+$0.05
 to correct for the updated electron-conduction opacities \citep{cas07}.
 This Figure shows that more than half of the RR~Lyrae stars are concentrated in
 the faintest $\sim$0.1 mag of the instability strip (IS), while the other half
 is spread over $\sim$0.3 mag above.
 From the evolutionary tracks, it appears that most of the variables belong to
 the fainter, more-metal rich sample; these stars tend to have shorter periods.
 The brighter variables, on the other hand, appear to be related either to the
 Z=0.0001 population, or to stars with slightly higher metallicity
 (Z$\sim$0.0003) that have evolved off the blue horizontal-branch visible on
 the WFPC2 CMD in Fig.~\ref{fig:1}.

 In any case, the range of luminosity covered by the HB stars within the IS
 cannot be explained by a monometallic population, even taking into account
 evolution off the ZAHB. Given that RR~Lyrae stars are old stars ($\ga$10~Gyr),
 this implies a relatively quick chemical enrichment in the first few billion
 years of the formation of the galaxy.

\subsection{WFPC2 sample}

 In the parallel WFPC2 field we discovered nine RR Lyrae stars (4 RR$ab$, 4
 RR$c$ and 1 RR$d$), even though the low signal-to-noise ratio of the
 observations did not allow us to measure their periods as accurately as for
 the variables in the ACS field-of-view.
 Based on the small samples of RR~Lyrae stars, we find
 $\langle$P$_{ab}\rangle$=0.59$\pm$0.03 and $\langle$P$_c\rangle$=0.35$\pm$0.01
 for the fundamental and first-overtone pulsators, respectively, which is in
 rough agreement with the values found for the RR~Lyrae stars in the ACS
 field-of-view.
 We provide their coordinates and approximate properties for completeness in
 Table~\ref{tab6}.

\begin{figure}
\epsscale{1.17} 
\plotone{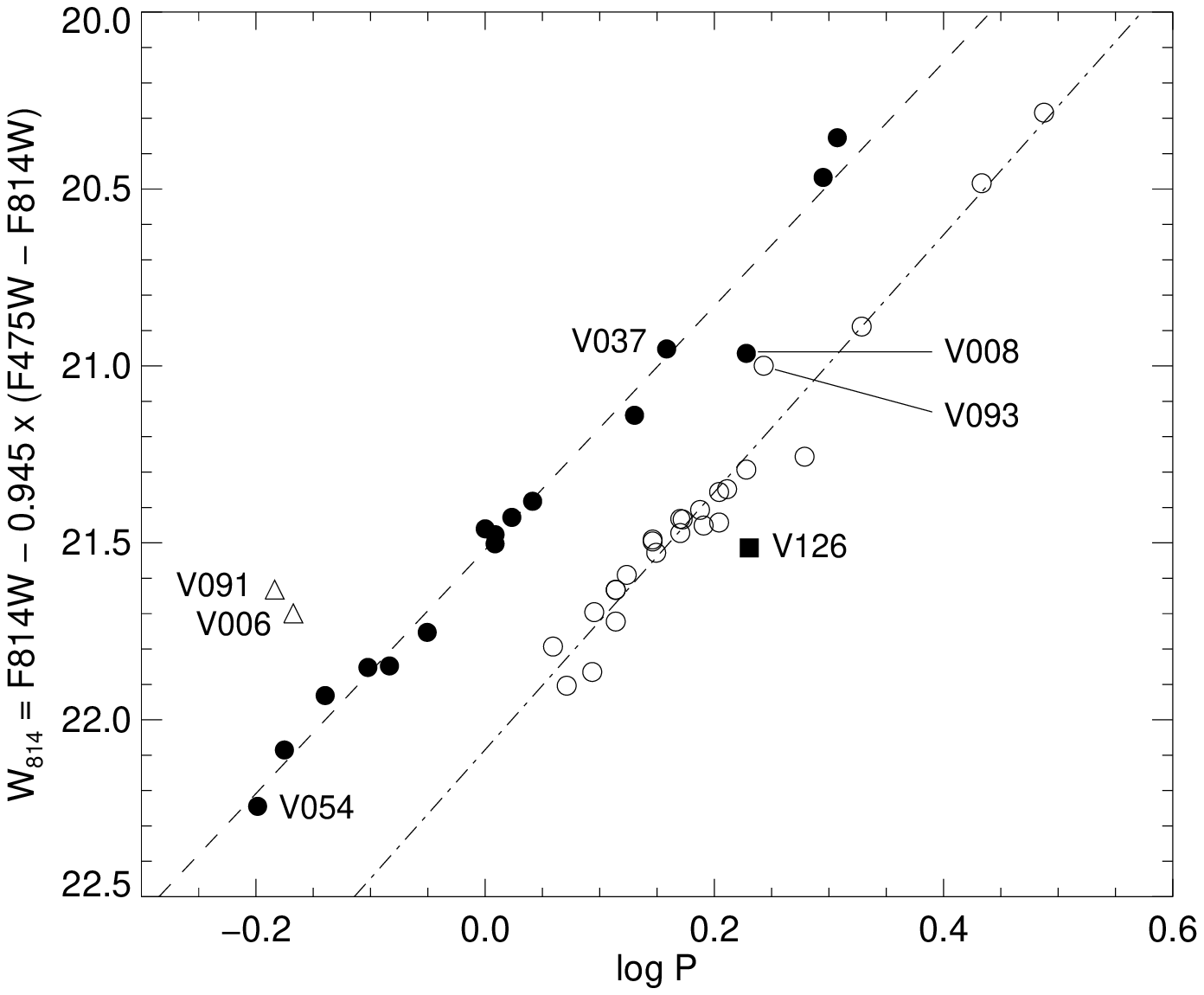}
\figcaption{Period-Wesenheit diagram for Cepheids in IC\,1613 ({\it see text
 for details}). The dash-dotted and dashed lines are linear fits to the
 fundamental and first-overtone Cepheids, respectively. The individual Cepheids
 mentioned in \S~\ref{sec:6.1}, \ref{sec:6.2}, \ref{sec:6.3}, and
 \ref{sec:11.1.1} are labeled.
\label{fig:9}}
\end{figure}

\begin{figure}
\epsscale{1.28} 
\plotone{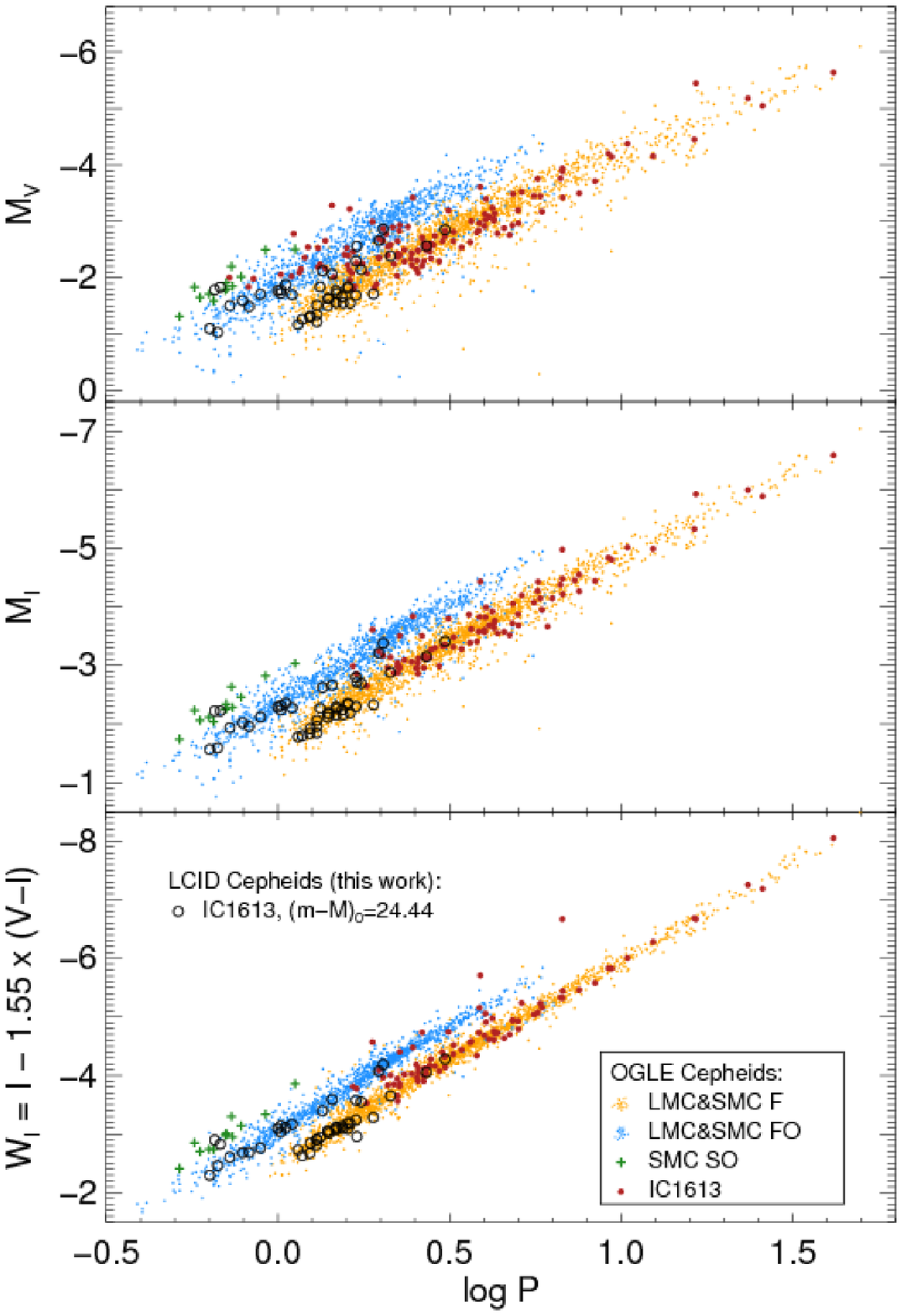}
\figcaption{Period-luminosity diagrams for Cepheids in IC\,1613.
 Our Cepheids are shown as open circles, while the symbols used for the OGLE
 Cepheids are as labeled in the inset.
\label{fig:10}}
\end{figure}

\subsection{Comparison with known RR~Lyrae stars in IC\,1613}\label{sec:5.3}

 Prior to our study, two surveys for variable stars found candidate RR~Lyrae
 stars in IC\,1613, although none of them in common with those presented here
 as the pointings did not overlap.
 Using the 4-shooter on the Hale-5m telescope, \citet{sah92} found 15 candidate
 RR~Lyrae stars in a 8$\arcmin \times$8$\arcmin$ field of view located
 12$\arcmin$ to the west of the galaxy center. From their Table~3, we
 calculated a mean period for the RR$ab$ of 0.60$\pm$0.02 day, which is in good
 agreement with our value. We note, however, that all their RR~Lyrae stars are
 of the $ab$ type and on average appear brighter by about 0.3 mag than expected
 from the distance and metallicity of IC\,1613 (see also \S~\ref{sec:11}).
 While some RR$c$ stars could have been missed due to their low amplitude and
 the relatively noisy light-curves, some of them might have been misidentified
 as RR$ab$ because of the small number of datapoints.
 Alternatively, given that crowding seems to be insignificant in this
 field---according to the finding charts---and that their RR$ab$ span
 a range of luminosity of over a magnitude, we suspect that some of their
 RR~Lyrae candidates could actually be short-period Cepheids.

\begin{figure}
\epsscale{1.2} 
\plotone{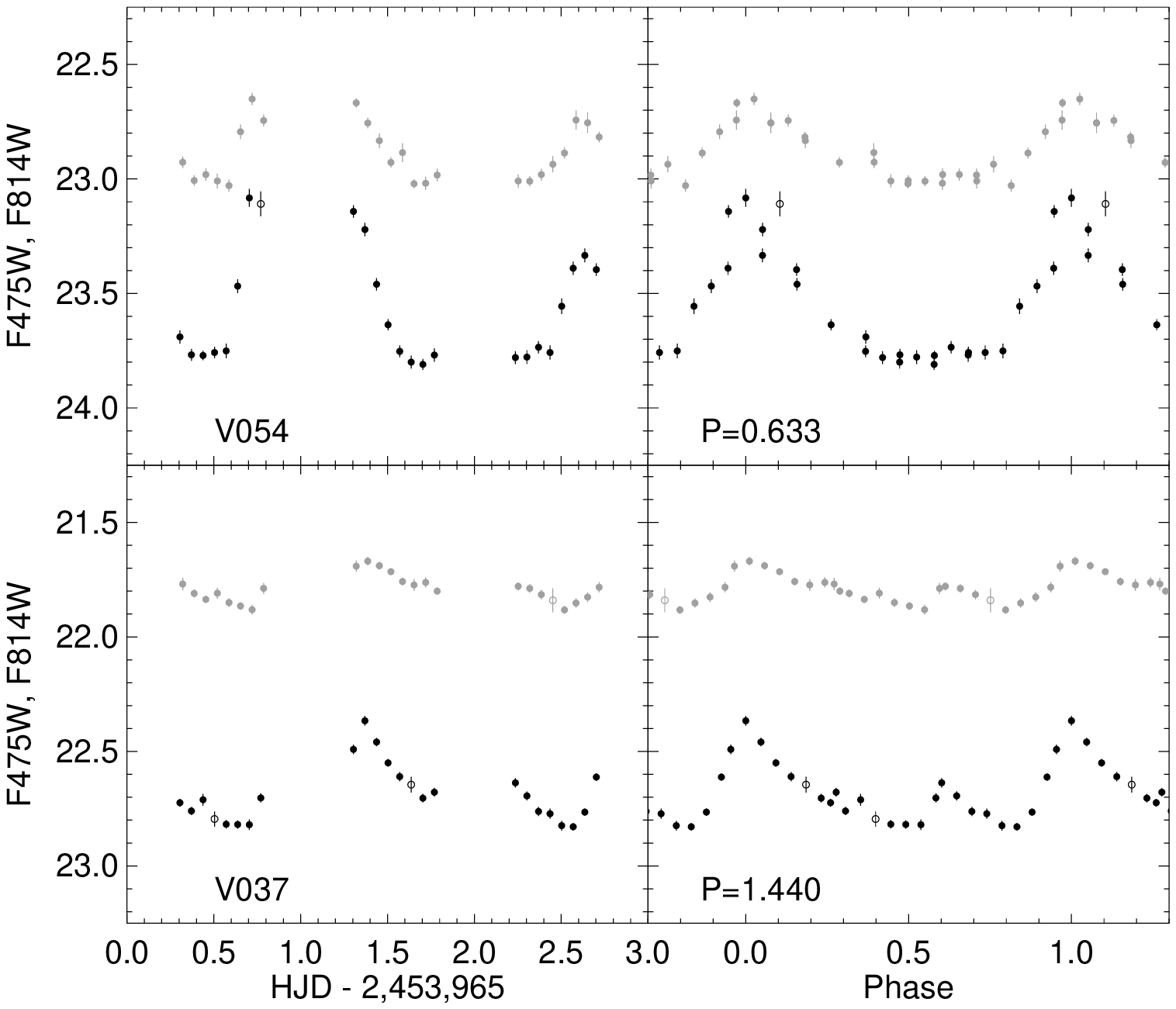}
\figcaption{The peculiar Cepheids V054 ({\it top}) and V037 ({\it bottom}),
 plotted as a function of Julian date ({\it left}) and phase ({\it right}).
\label{fig:11}}
\end{figure}

 The analysis of WFPC2 data by \citet{dol01} in a field 7.4$\arcmin$ southwest
 from the center, on the other hand, returned 13 RR~Lyrae candidates. Assuming
 that the likely and possible overtone pulsators are actual RR$c$, their field
 contains 9 RR$ab$ and 4 RR$c$ with mean periods of $\sim$0.58$\pm$0.02 and
 $\sim$0.37$\pm$0.03 day, respectively.
 This is very similar to the values we found for our WFPC2 field. The slightly
 shorter mean period for the RR$ab$, when compared to our ACS sample, may be
 due to the difficulty of detecting the low amplitude, longer period variables
 because of the lower quality of the data.

 In any case, these samples at larger radii are not large enough for a robust
 comparison with the ACS sample in order to reveal potential gradients in the
 properties of the RR~Lyrae stars as was done for the Tucana dwarf in
 \citet{ber08}. In the case of Tucana, the ACS field of view covered a larger
 fraction of the radius of the galaxy, allowing a radial gradient study within
 the ACS field alone.

\section{Cepheids}\label{sec:6}

 In the search for variable stars presented here we found 44 variables with
 properties of Cepheids: about 2 mag brighter than the HB, in or close to the
 IS, and P$\ga$0.6 days. Their light-curves are shown in Fig.~\ref{fig:8}.
 Another five were found for which we could not determine their period and/or
 intensity-averaged mean magnitude due to inadequate temporal sampling. These
 are discussed in \S~\ref{sec:6.4}.

 Figure~\ref{fig:9} shows the period$-W_{814}$ diagram for the Cepheids in
 IC\,1613. Because Cepheids from the different modes of pulsation have a
 different magnitude at a given period, their location in the period-luminosity
 (PL) diagrams can be used to differentiate their type. These PL
 relations are fundamental in that they represent the most robust method to
 calculate the distance to galaxies within the nearby universe.
 To reduce the scatter due to interstellar extinction and the intrinsic
 dispersion due to the finite width of the instability-strip, and to obtain a
 more secure classification, we calculated the Wesenheit, or reddening-free,
 magnitude introduced by \citet{van68}. In the {\it HST} bands used here, the
 relation is of the form:
\begin{equation}
 W_{814} = F814W - 0.945 (F475W-F814W)
\end{equation}
 where the coefficient 0.945 comes from the standard interstellar extinction
 curve dependence of the F814W magnitude on E(F475W$-$F814W), from
 \citet{sch98}.

 Figure~\ref{fig:9} shows two almost parallel sequences, and correspond to the
 fundamental (F, {\it open circles}) and first-overtone (FO, {\it filled
 circles}) Cepheids, respectively. A few outliers are marked as open triangles
 (V006 and V091) and a filled square (V126), and are discussed in
 \S~\ref{sec:6.2} and \ref{sec:6.3}, respectively. From this plot, we find that
 25 Cepheids are pulsating in the fundamental mode, while 16 fall on the
 first-overtone sequence.

 Figure~\ref{fig:10} shows the PL diagram for the Johnson VI bands obtained as
 described in \citetalias{ber09}, as well as in the W$_I$ band as defined in
 \citet[$W_I = I - 1.55 (V-I)$]{uda99c}. The new Cepheids of IC\,1613
 ({\it open circles}) are shown overplotted on the classical Cepheids of the
 Large and Small Magellanic Clouds (LMC \& SMC) and of IC\,1613 from the OGLE
 collaboration \citep[and references therein]{uda01} as labeled in the inset.
 The apparent magnitudes were converted to absolute magnitude assuming a
 distance modulus of 18.515$\pm$0.085 to the LMC \citep{cle03}, and a distance
 offset of 0.51 of the SMC relative to the LMC \citep{uda99c}. The Cepheids of
 IC\,1613 were shifted according to the distance determined in \S~\ref{sec:11}.
 It shows that the properties of the Cepheids discovered in our field are in
 excellent agreement with the properties of OGLE Cepheids, therefore confirming
 their classical Cepheid nature.

 However, one can see that the Cepheids in IC\,1613 (both ours and OGLE) are
 mainly located toward the short-period extremity of the fundamental and
 overtone sequences when compared to the Cepheids in the Magellanic Clouds.
 Studies of the variable stars in the very metal poor dwarf irregular galaxies
 Leo~A \citep{dol02} and Sextans~A \citep{dol03} also revealed large numbers of
 short period Cepheids. This agrees with the suggestion by \citet{bon03} that
 ``the minimum mass that performs the blue loop [decreases with decreasing
 metallicity, which] means that metal-poor stellar systems such as IC\,1613
 should produce a substantial fraction of short-period classical Cepheids.''

 Interestingly, the IC\,1613 fundamental-mode Cepheids perfectly follow the
 various PL relationships defined by the Magellanic Clouds Cepheids over the
 whole range of periods, from $\sim$1 to $\sim$40 days (see Fig.~\ref{fig:10}).
 This suggests that the metallicity dependence of these relations, if any, must
 be very small at such low metallicities.

\subsection{The peculiar Cepheids V037 \& V054}\label{sec:6.1}

 Two of the Cepheids falling on the first-overtone sequence in Fig.~\ref{fig:9}
 present very peculiar light-curves. These are shown again in Fig.~\ref{fig:11},
 plotted as a function of Julian date ({\it left}) and phase ({\it right}).

 In the top left panel, the amplitude of V054 seems to be decreasing at each
 consecutive maximum. The dispersion at maximum light on the phased
 light-curves is reminiscent of the light-curves of RR$d$ stars, although in
 this particular case the periodogram does not present the double-peak features
 characteristic of double-mode stars. However, the rather long main period of
 this variable (0.633 day vs. $\sim$0.39 for RR$d$) limited the number of
 observed cycles to $\la$4 and might explain the lack of resolution in the
 periodogram.
 In addition, the position of V054 on the PL diagrams, at the shorter period
 end, is in excellent agreement with the double-mode Cepheids pulsating
 simultaneously in the fundamental and first-overtone mode presented in
 \citet{sos08a}.

 In the case of V037, we found that the only period giving a smooth light-curve
 is P=1.44 days. However, as shown in the bottom right panel of
 Fig.~\ref{fig:11}, this produces a light-curve with a secondary maximum at
 phase $\sim$0.6. If this period is confirmed to be true, this would be the
 second Cepheid with such a light-curve in IC\,1613 after V39 discovered by
 Baade \citep[see][]{san71}. Note, however, that more recent observations and
 analysis of V39 \citep[e.g.,][]{man01} favored a period only half that given
 by \citet[P=14.350 vs. 28.72 days]{san71} associated with a long term
 modulation of $\sim$1100 days, which could correspond to an extremely distant
 W~Vir star in the outer halo of the MW blended with a long-period variable in
 IC\,1613.

 In our case, the short timebase of our observations definitely rules out the
 possibility of long term modulation. In addition, a shorter period for V037
 cannot phase the light-curve properly unless a strong modulation of the period
 itself is taken into account, which is very unlikely on such short timescales.
 Given that its position in the CMD and PL diagrams is exactly that expected
 from a Cepheid, in the following we will assume that V037 is a bona fide
 classical Cepheid.

\begin{figure}
\epsscale{1.29} 
\plotone{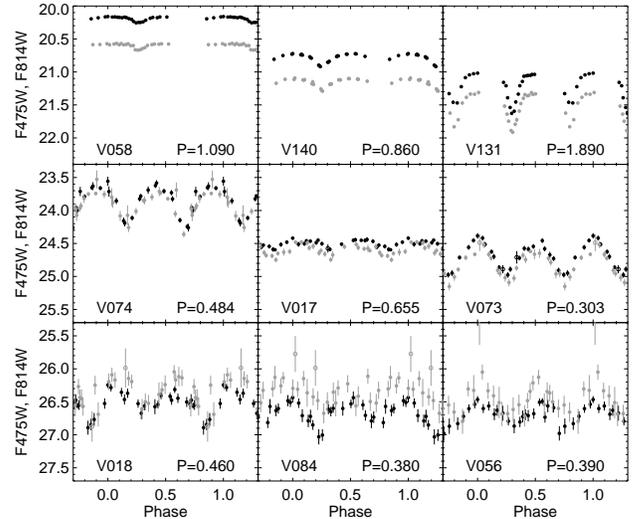}
\figcaption{Sample light-curves for eclipsing binaries in IC\,1613. The three
 brightest ({\it top}), three intermediate magnitude ({\it middle}), and the
 three faintest ({\it bottom}) binaries are shown.
\label{fig:12}}
\end{figure}

\subsection{Second-Overtone candidates V006 \& V091}\label{sec:6.2}

 In the diagrams shown in Figs.~\ref{fig:9} \& \ref{fig:10}, two Cepheids lie
 above the first-overtone sequence in all panels. In particular in
 Fig.~\ref{fig:10}, they follow the sequence representing the second-overtone
 pulsators found by the OGLE collaboration in the SMC. Their nearly sinusoidal
 light-curve with very low amplitude in both bands ($\la$0.14 in F475W,
 $\sim$0.07 in F814W), as well as their location well within the IS, make them
 very strong second-overtone candidates. These are the first ones detected
 beyond the Magellanic Clouds.

 Second-overtone classical Cepheids are very rare objects, as witnessed by the
 few firm candidates known to date: 13 out of 2049 classical
 Cepheids---0.42\%---in the SMC \citep{uda99a}, and 14 out of
 3361---0.63\%---in the LMC \citep{sos08a}. However, theoretical models by
 \citet{ant97} showed that these Cepheids are more likely to be found in low
 metallicity systems, which is in qualitative agreement with the fraction of
 second-overtone Cepheids in our IC\,1613 sample, about 4\%.

\subsection{V126: a blended Cepheid?}\label{sec:6.3}

 The light-curve morphology, luminosity and period of V126 are all
 characteristic of a classical Cepheid. On the other hand, its color is 0.5 mag
 bluer than the blue edge of the IS, and therefore appears below the
 fundamental mode pulsator sequence in the period-Wesenheit diagram shown in
 Fig.~\ref{fig:9}.
 Although it appears isolated on the stacked images, the most likely hypothesis
 is that it is a bona fide Cepheid blended with an unresolved main-sequence
 star. The low amplitude of its luminosity variation in both bands relative
 to the other Cepheids with similar periods seems to support this conclusion.

\subsection{Other candidates}\label{sec:6.4}

 Five Cepheids were observed for which we could not accurately determine their
 period and/or mean magnitude due to inadequate temporal sampling. We use the
 weighted mean magnitude given by ALLFRAME as their approximate magnitudes.
 They are shown as open stars in Fig.~\ref{fig:1}.

 V030 has a period $\sim$0.98 day, and was therefore observed at the same
 phase each day, at minimum light. Most of the rising phase and peak are
 missing so it was not possible to fit a template light-curve and measure
 accurately its mean magnitude. From its approximate magnitude, its
 location on the PL diagram is consistent with the fundamental mode Cepheids.

 V090 has relatively low-amplitude (A$_{475}$$\sim$0.15, A$_{814}$$\sim$0.11)
 and period 2$\la$P$\la$4 days. From its location on the PL diagram it is a
 probable first-overtone.

 V124 is very bright and was previously known from ground-based observations
 (see \S~\ref{sec:10}). Given its very long period ($\sim$26 days), our
 observations cover less than 10\% of a cycle so we did not try to obtain the
 mean magnitude from our data. It is shown in Fig.~\ref{fig:10} as the
 fundamental mode OGLE Cepheid at log P$=$1.41.

 V137 was also present in the OGLE field, although their quoted period of
 1.376~days does not provide a satisfactory fit to our data. Even though we
 could not obtain an accurate period for this variable, the 1.4-mag decrease in
 the F475W band over the first two days and the beginning of the third, and the
 subsequent increase implies a period in the range 2.7$\la$P$\la$3.3 days, which
 would make it a fundamental mode Cepheid.

 V151 is one of the brightest Cepheids of our sample, and thus has a period
 longer than our observational timebase. Unfortunately, it is missing from the
 OGLE catalog so we can only constrain the period to the range
 2.5$\la$P$\la$3.1 days from the light-curve morphology. It places it on the PL
 relation of fundamental mode Cepheids.

\section{Eclipsing Binary Stars}\label{sec:7}

 A large number of variable star candidates were found in the main-sequence,
 as can be expected from the fact that it is well populated from the turn-offs
 to the B giants. Most of the variables for which we could find a period are
 eclipsing binary stars. The remaining ones have properties of high-amplitude
 $\delta$-Scuti stars (HADS) and are discussed in the following section, while
 the candidates for which the light-curves could not be phased are presented in
 \S~\ref{sec:9}. Note that all of these stars are newly discovered variables.

 In the ACS field-of-view, we found 38 eclipsing binaries between `V'$\sim$20.5
 and $\sim$26.5. At fainter magnitudes, the signal-to-noise ratio of
 the individual datapoints was too low to estimate the period and/or confirm
 variability. The light-curves of the three brightest, three intermediate
 magnitude, and the three faintest binaries are shown in Fig.~\ref{fig:12}.

 Given the small number of datapoints and relatively low SNR of most of the
 candidates, some periods might actually be multiples of the true periods. We
 thus did not try to assign them a particular type of binary (detached or
 contact)---although the variety of light-curve morphologies indicate that
 members of both kinds appear to be present---nor check for their membership in
 IC\,1613.

 Bright eclipsing binaries have been used to determine distances to LG galaxies
 with claimed accuracy better than about 5\% (e.g., M31: \citealt{rib05};
 M33: \citealt{bon06}). Unfortunately, the stars need to be bright enough
 (V$_{max}<$20) to accomplish this in a reasonable time with current observing
 facilities. Even though some of our eclipsing binaries have accurate periods
 and deep eclipses, none of them is brighter than V$\sim$20.4.

\begin{figure}
\epsscale{1.1} 
\plotone{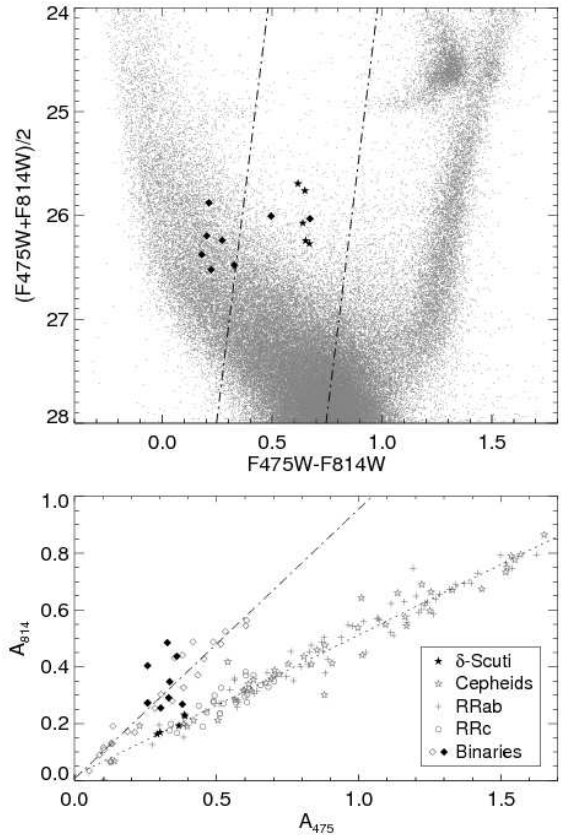}
\figcaption{Selection criteria for the $\delta$-Scuti candidates.
 {\it Top:} Detail of the CMD, where the possible candidates are overplotted
 with larger symbols. The dash-dotted lines indicate the approximate location
 of the classical instability strip fitted on these data.
 {\it Bottom:} F814W versus F475W amplitudes, where the pulsating variables and
 binary stars have a different slope. Pluses, open circles, and open stars are
 for the RR$ab$, RR$c$, and Cepheids, respectively, while the open diamonds
 represent the binaries with good light-curves. The dotted and dash-dotted
 lines are linear fits to the pulsating variables and binaries, respectively.
 In both panels the $\delta$-Scuti candidates satisfying both requirements are
 shown as filled stars, and the rejected candidates as filled diamonds.
\label{fig:13}}
\end{figure}

\section{Candidate $\delta$-Scuti stars}\label{sec:8}

 Among the candidate variables fainter than the HB that appeared to be periodic
 variables, most of them were found to be binaries and are described in the
 previous section. For some of these variables, however, it was possible to
 obtain a relatively smooth, pulsation-like light-curve (i.e., with a single
 minimum) with a short period. This was the case for 13 variables, which were
 flagged as possible $\delta$-Scuti stars. To further constrain their nature,
 additional parameters were taken into account.
 First, we required that their color be within the classical IS, defined
 approximately by the position of the Cepheids and RR~Lyrae stars. The top
 panel of Fig.~\ref{fig:13} shows a zoom-in of the CMD showing the location of
 the IS, where the 13 possible $\delta$-Scuti are shown as filled symbols.
 Eight of these fall within or close to the boundaries of the IS.

 Given that the pulsations of the $\delta$-Scuti stars, like the Cepheids and
 RR~Lyrae stars are driven by the $\kappa$-mechanism, the second criteria was
 that the amplitude ratio had to be in agreement with that of the other
 pulsating variables. The bottom panel of Fig.~\ref{fig:13} shows the amplitude
 in F814W versus the amplitude in F475W for Cepheids, RR~Lyrae stars, eclipsing
 binaries with good light-curves, and the possible $\delta$-Scuti stars.
 Only 5 satisfy this requirement, and are shown as filled stars in both panels.
 They also present relatively smooth light-curves with very short periods
 ($\la$0.2 days). Their pulsational properties are summarized in
 Table~\ref{tab4}.

\begin{figure}
\epsscale{1.28} 
\plotone{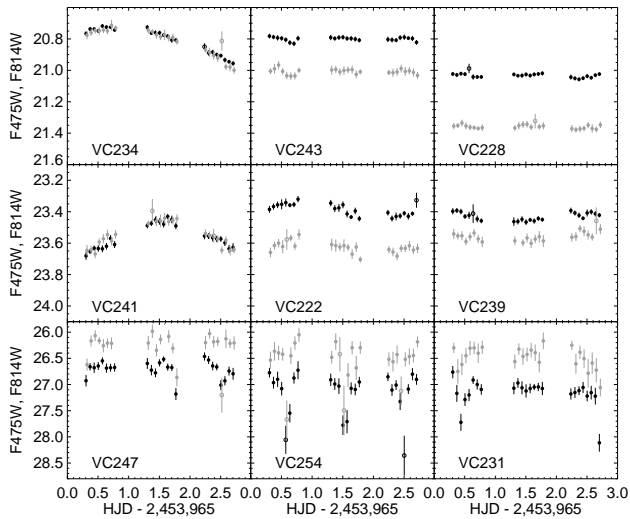}
\figcaption{Sample light-curves for MSVs in IC\,1613, as a function of HJD. As
 in Fig.~\ref{fig:12}, the three brightest ({\it top}), three intermediate
 magnitude ({\it middle}), and the three faintest ({\it bottom}) MSV candidates
 are shown. Note the different vertical scale for the bottom panels.
\label{fig:14}}
\end{figure}

\section{Other Variables}\label{sec:9}

 In addition to the classical IS variables and eclipsing binaries presented
 above, we detected another 77 candidate variables throughout the CMD, for
 which we could not determine the period because of inadequate temporal
 sampling. Most of these candidate variables are located on the MS, although
 five LPV candidates are also present at or close to the TRGB.
 While the majority of the MS candidates are most likely eclipsing binaries, at
 least in the brightest part of the MS might there be pulsating variables of
 the $\beta$ Cephei or Slowly Pulsating B stars types
 \citep[see, e.g.,][]{dia08}. A sample of the brightest, intermediate, and the
 faintest light-curves is shown in Fig.~\ref{fig:14}.

\section{Previously Known Variables}\label{sec:10}

 The variable stars in our ACS field in IC\,1613 that were already known prior
 to this work are presented in Table~\ref{tab7}. We give their index and period
 as measured from our data, as well as their identifications and periods in
 previous catalogs; most of them were discovered during the wide-field survey by
 the OGLE collaboration. Because of the relatively shallow magnitude limits of
 the other surveys, all are luminous Cepheids with relatively large amplitude.

 Two of them, V124 and V172, were already included in Baade's catalog
 \citep{san71} as V11 and V35, respectively. These are the two brightest
 Cepheids of our sample, although the very long period of the former prevented
 us from measuring its intensity-averaged magnitude and period. Another
 candidate variable that was flagged as a possible Cepheid in \citet{san71} and
 judged to be an irregular variable in \citet{car90}, V41 is constant within
 $\sim$0.02 and 0.05 mag in our F475W and F814W data and was not included in our
 catalog.

 Two other Cepheids appeared in \citet{man01} with periods similar to the ones
 found here within 0.01 day, while three LPV/Irr variables of their catalog
 appear constant in our data (V0762C, V1598C, and V1667C). However, all have
 very long periods ($\ga$70 days) and/or low amplitude (A$_{Wh}$=0.20) and their
 variation could have easily been missed in our data due to the short observing
 run.

 The remaining variables were previously observed by OGLE \citep{uda01}. For
 most of them, the periods determined from our data agree well with theirs.
 However, the 1.376 day period given by OGLE for V137 cannot phase our data.
 Our sampling over 3 consecutive days clearly covers only about one cycle, so
 the period must be very close to P=3 days.

 By comparing the number of Cepheids observed by us and the OGLE team in the
 same field, we can obtain a rough estimate of the completeness of their
 survey. We found 49 Cepheids over the area of the ACS camera, while their
 sensitivity limited them to 12 detections.
 Assuming that the density of Cepheids is uniform in the central region of
 IC\,1613, this means that about 500 Cepheids should be present in the
 14$\arcmin$x14$\arcmin$ field observed by the OGLE collaboration. Given the
 extent of the galaxy, this value is roughly in agreement with the total
 number estimated in \S~\ref{sec:4}.


\begin{deluxetable}{rccccc}
\tablewidth{0pt}
\tablecaption{Cross-Identification with Previously Known Cepheids in IC\,1613\label{tab7}}
\tablehead{
\colhead{ID} & \colhead{Period} & \colhead{ID$_{OGLE}$} &
\colhead{P$_{OGLE}$} & \colhead{ID$_{other}$\tablenotemark{a}} &
\colhead{P$_{other}$}}
\startdata
  V008 &       1.69 & 2124 &  1.697 & V1478C &  1.701    \\
  V022 &       1.48 & 2197 &  1.459 &    -   &    -      \\
  V057 &       1.69 & 2818 &  1.661 &    -   &    -      \\
  V077 &       1.35 & 2909 &  1.309 &    -   &    -      \\
  V086 &        -   & 2389 &  2.029 &    -   &    -      \\
  V093 &       1.75 &   -  &    -   & V1465C &  1.741    \\
  V106 &       2.13 & 2117 &  2.131 &    -   &    -      \\
  V124 &        -   & 1987 & 25.862 &   V11  & 25.7719   \\
  V137 & $\sim$3.   & 2751 &  1.376 &    -   &    -      \\
  V147 &       2.0  & 2342 &  1.972 &    -   &    -      \\
  V154 &       2.7  & 2760 &  2.712 &    -   &    -      \\
  V170 &       1.33 & 2771 &  1.329 &    -   &    -      \\
  V172 &       3.0  & 2240 &  3.074 &   V35  &  3.073417
\enddata
\tablenotetext{a}{V11 \& V35 from \citet{san71}, V1478C \& V1465C from
 \citet{man01}}
\end{deluxetable}


\section{Distance}\label{sec:11}

 In \citetalias{ber09}, we used the photometric and pulsational properties of
 the RR~Lyrae stars to measure the distances to the dSphs Cetus and Tucana, and
 showed that the values we obtained are in excellent agreement with the
 distances obtained previously with independent methods. We therefore obtain
 here a RR~Lyrae-based distance for IC\,1613 in the same way. Additionally,
 given the significant number of Cepheids we found in IC\,1613 and the high
 quality of their light-curves, together with their firm classification as
 fundamental, {first-,} or second-overtone pulsators, it is possible to
 calculate accurate distances from their properties as well.

 In this section, we use several methods adopted in the literature to
 calculate the distance based on the photometric and pulsational properties of
 the Cepheids and RR~Lyrae stars. In all cases, the intensity-averaged mean
 magnitudes in the Johnson bands are used.

\subsection{Cepheids}\label{sec:11.1}

\subsubsection{Period-Wesenheit relation for Fundamental and First-Overtone Cepheids}\label{sec:11.1.1}

 The main distance indicator for star forming galaxies in the nearby universe
 is the relation between the period and the luminosity of their Cepheids, even
 though the slope and zero-point of this relation might be slightly dependent
 on metallicity \citep[e.g.,][]{ken98,sak04,san08}. However, from the analysis
 of nonlinear convective pulsations models, \citet{fio07} showed that the I-band
 period-luminosity is not expected to show metallicity dependence at
 metallicities lower than that of the LMC (Z=0.008), and that the effect of
 metal-content on the P--M$_V$ relation is negligible below Z=0.004. Given the
 low metallicity of the young stars in IC\,1613 (Z$\sim$0.003), we will
 therefore assume that it is safe to estimate its distance by comparison with
 the properties of the Cepheids of the Magellanic Clouds.

 Here we chose to derive a new P--$W_I$ (see \S~\ref{sec:6}) relation
 based on the Cepheids of both MCs instead of using the relations available in
 the literature for the following reasons:
 i) first, the domain of validity for the PL relations of the literature starts
 at about P=2.5 days \citep[log P = 0.4; see e.g.,][]{uda99c,fou07}, while all
 our well-measured Cepheids have periods shorter than this value; and
 ii) the period distribution of the Cepheids in the LMC and SMC are very
 different from each other, with the peaks in the distribution at P$\sim$3.2 and
 2.1 days for the fundamental and first-overtone Cepheids in the LMC, and at
 P$\sim$1.6 and 1.3 days, respectively, in the SMC. While the period
 distribution of the Cepheids in IC\,1613 (with the peaks at P$\sim$1.6 and 1.1
 days) are very similar to those in the SMC, the distances to other galaxies are
 usually provided relative to a given LMC distance.
 In addition, the PL relation of the SMC presents a rather large dispersion due
 to the inclination of the galaxy with respect to the line of sight.
 We thus combined the Cepheids of both Magellanic Clouds assuming
 $\Delta$(m$-$M)$_0$=0.51$\pm$0.03 \citep{uda99c} to improve the period
 coverage. This also has the advantage of cancelling any small difference in
 slope that might be present between the PL relations of the LMC and SMC.

 In addition, most of the Cepheids we observed in IC\,1613 are located at the
 short-period end of the distribution, so any non-linearity in the PL relation
 would lead to a different distance had the Cepheids had longer periods.
 \citet{nge08} showed that the PL relation in the Wesenheit magnitude $W_I$
 is linear over the period range covered by the LMC Cepheids, while it is not
 in the V\&I bands.
 Finally, the use of $W_I$ also limits the effect of interstellar reddening
 and thus reduces the scatter in the relation.

 We calculate the P--$W_I$ relation from linear regression fits to the combined
 MC Cepheids from OGLE-II observations \citep{uda99b,uda99d} over the whole
 range of periods after rejecting the outliers through sigma-clipping
 (2.5-$\sigma$, 5 iterations). Assuming an LMC distance of
 (m$-$M)$_0$=18.515$\pm$0.085 \citep{cle03}, we found:

     W$_{I,F}$=$-$3.435($\pm$0.007) log $P_F$ $-$ 2.540($\pm$0.006), and

     W$_{I,FO}$=$-$3.544($\pm$0.013) log $P_{FO}$ $-$ 3.067($\pm$0.007)

 \noindent for the fundamental and first-overtone Cepheids, respectively, with
 a standard deviation of 0.096 in both cases.

 The slope of our fundamental P--$W_I$ relation is marginally different from
 the slope found by other authors (e.g., W$_{I,F}$/log\,$P_F$=3.313$\pm$0.008,
 \citealt{nge08}; W$_{I,F}$/log\,$P_F$=3.320$\pm$0.011, \citealt{fou07})
 based on the OGLE Cepheids of the LMC only, and may be an indication that
 this relation deviates from linearity at periods shorter than about 0.3 days.
 On the other hand, it is intermediate between the slopes found for the LMC and
 the values obtained for the Cepheids of the MW
 \citep[e.g.,][W$_{I,F}$/log\,$P_F$=3.477$\pm$0.074]{fou07}.

\begin{figure}
\epsscale{1.25} 
\plotone{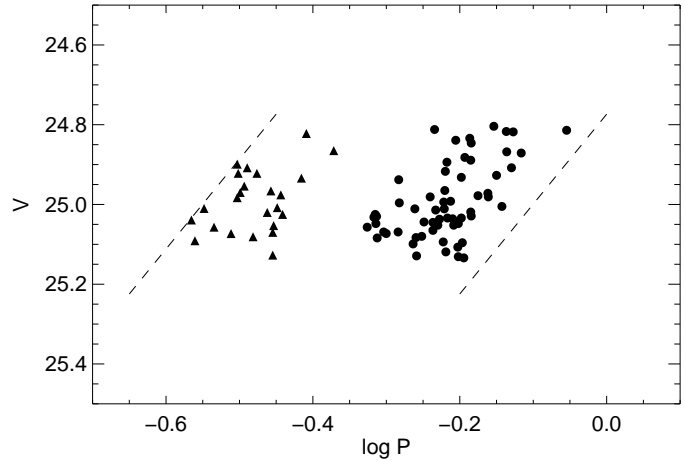}
\figcaption{Distribution of RR Lyrae in the $\langle V \rangle$-log P plane,
 where the predicted edges of the instability strip of \citet{cap00} have been
 overplotted ({\it see text for details}). Filled circles and triangles
 represent RR$ab$ and RR$c$, respectively.
\label{fig:15}}
\end{figure}

 We applied the P--$W_I$ relations derived above to the fundamental and
 first-overtone Cepheids of IC\,1613 shown in Fig.~\ref{fig:9} (excluding the
 two Cepheids located at the midpoint between the dash-dotted and dashed lines,
 V008 and V093), and found (m$-$M)$_0$=24.50$\pm$0.11 and 24.47$\pm$0.12,
 respectively. The good agreement between these two values indicates that
 first-overtone Cepheids can be safely used as distance indicators in the case
 where the pulsating mode of the Cepheids can be unambiguously identified.

 We also calculated the distance to IC\,1613 using the PL relation for
 fundamental-mode Cepheids given by \citet{nge08} to check if the small
 differences in slope and zero-point described above would have a significant
 effect, but found a very similar value ((m$-$M)$_0$=24.55$\pm$0.12) within the
 uncertainties.


\begin{deluxetable}{llcclc}
\tabletypesize{\scriptsize}
\tablewidth{0pt}
\tablecaption{Summary of distance moduli to IC\,1613.\label{tab8}}
\tablehead{
\colhead{(m$-$M)} & \colhead{E(B$-$V)} & \colhead{$\mu_{0,LMC}$} &
\colhead{Band} & \colhead{Method} & \colhead{Ref.}}
\startdata
   24.50$\pm$0.11  &  0.025 & 18.515 &   W$_I$   & Cep: F PL         & 1   \\
   24.47$\pm$0.12  &  0.025 & 18.515 &   W$_I$   & Cep: FO PL        & 1   \\
   24.46$\pm$0.11  &  0.025 & 18.515 &     VI    & Cep: SO PLC       & 1   \\
   24.47$\pm$0.12  &  0.025 & 18.515 &     V     & RRL: M$_V$-[Fe/H] & 1   \\
   24.44$\pm$0.10  &  0.025 & 18.515 &     V     & RRL: PLM          & 1   \\
\hline
   24.55           &  0.03  &  $-$   &     B     & Cep: F PL         & 2   \\
   24.28$\pm$0.25  &  0.05  & 18.27  &     B     & Cep: F PL         & 3   \\
   24.32$\pm$0.11  &  0.03  &  $-$   &     H     & Cep: F PL         & 4   \\
   24.27$\pm$0.11  &  0.04  & 18.50  &   BVRIH   & Cep: F PL         & 5   \\
   24.39$\pm$0.14  &  0.04  & 18.50  &   W$_I$   & Cep: F PL         & 5   \\
   24.17$\pm$0.27  &  0.02  &  $-$   &     g     & RRL: M$_V$-[Fe/H] & 6   \\
   24.53$\pm$0.13  &  0.03  & 18.50  &     V     & Cep: F PL         & 7   \\
   24.44$\pm$0.13  &  0.03  & 18.50  &     I     & Cep: F PL         & 7   \\
   24.53$\pm$0.12  &  0.03  & 18.50  &     J     & Cep: F PL         & 7   \\
   24.43$\pm$0.08  &  0.03  & 18.50  &     H     & Cep: F PL         & 7   \\
   24.40$\pm$0.09  &  0.025 &  $-$   &     I     & TRGB              & 8   \\
   24.38$\pm$0.09  &  0.025 &  $-$   &     I     & Red Clump         & 8   \\
   24.40$\pm$0.16  &  0.025 &  $-$   &     V     & RRL: M$_V$-[Fe/H] & 8   \\
   24.31$\pm$0.07  &  0.025 & 18.21  &     V     & Cep: F PL         & 9   \\
   24.24$\pm$0.07  &  0.025 & 18.23  &     I     & Cep: F PL         & 9   \\
   24.24$\pm$0.07  &  0.025 & 18.25  &   W$_I$   & Cep: F PL         & 9   \\
   24.53$\pm$0.10  &  0.02  &  $-$   &     I     & TRGB              & 10  \\
   24.385          &  0.09  & 18.50  &     J     & Cep: F PL         & 11  \\
   24.306          &  0.09  & 18.50  &     K     & Cep: F PL         & 11  \\
   24.50$\pm$0.12  &  0.024 & 18.54  &   BVRI    & Cep: F PL         & 12  \\
   24.43$\pm$0.05  &  0.025 &  $-$   &     I     & TRGB              & 13  \\
   24.29$\pm$0.07  &  0.08  & 18.50  & 3.6$\mu$m & Cep: F PL         & 14  \\
   24.28$\pm$0.07  &  0.08  & 18.50  & 4.5$\mu$m & Cep: F PL         & 14
\enddata
\tablerefs{
(1) This work;
(2) \citealt{san71};
(3) \citealt{dev78};
(4) \citealt{mca84};
(5) \citealt{fre88};
(6) \citealt{sah92};
(7) \citealt{mac01};
(8) \citealt{dol01};
(9) \citealt{uda01};
(10) \citealt{tik02};
(11) \citealt{pie06};
(12) \citealt{ant06};
(13) \citealt{riz07};
(14) \citealt{fre09}.}
\end{deluxetable}


\subsubsection{Second-Overtone Cepheids and the Period-Luminosity-Color relation}

 From their theoretical work on second-overtone (SO) Cepheids, \citet{bon01}
 noted that these variables follow a period-luminosity-color relation of the
 form:
\begin{displaymath}
 M_V  =  - 3.961 - 3.905 log  P  + 3.250 (V - I).
\end{displaymath}
 Given the low amplitude of the luminosity variations and the very small
 temperature width of the SO IS, the standard deviation of this relation is
 only 0.004. It is thus possible to use it to determine distances with good
 accuracy. From the SO Cepheids observed in the SMC by \citet{uda99a} and
 assuming E(B$-$V)=0.054, the authors found a distance of
 (m$-$M)$_{0,SMC}$=19.11$\pm$0.08 to the SMC through this relation.

 Applying this relation to the two SO Cepheids described in \S~\ref{sec:6.2}, we
 find a distance modulus of (m$-$M)$_0$=24.54$\pm$0.11.
 Note, however, that the distance to the SMC used in the previous section to
 obtain the period-W$_I$ relationship was 19.03, assuming an LMC distance of
 18.515$\pm$0.085 and a distance moduli difference of 0.51$\pm$0.03. Taking
 into account this offset in zero-point, we find that the true distance modulus
 to IC\,1613 from SO Cepheids is 24.46$\pm$0.11, in better agreement with the
 mean value determined below.


\begin{figure}
\epsscale{1.2} 
\plotone{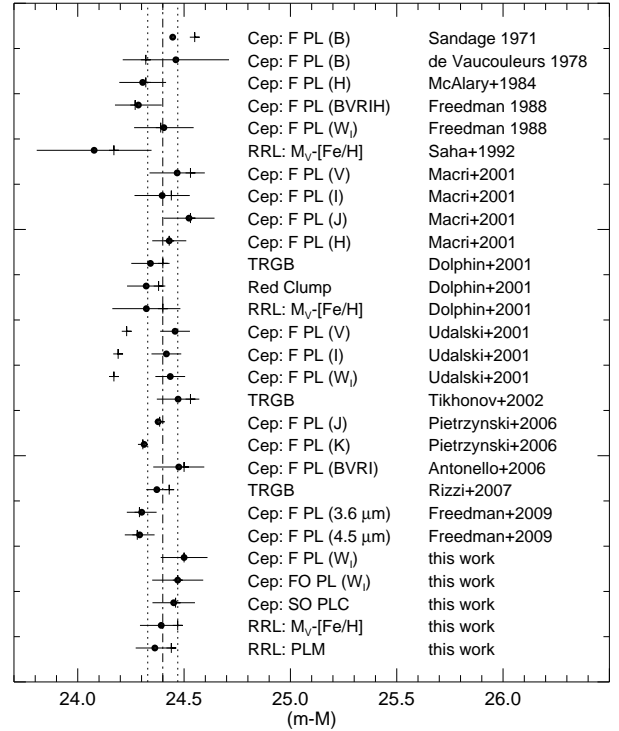}
\figcaption{Summary of the distance moduli from the present work and the
 literature. The reddened moduli as given in the referenced articles are shown
 as pluses, while the moduli corrected for a common reddening of E(B$-$V)=0.025
 and LMC distance modulus (m$-$M)$_0$=18.515$\pm$0.085 are shown as filled
 circles. The vertical lines indicate the mean and standard deviation
 ($\sigma$=0.071) of these measurements (excluding the \citet{sah92} outlier)
 at (m$-$M)$_0$=24.400$\pm$0.014 (statistical).
\label{fig:16}}
\end{figure}

\subsection{RR~Lyrae stars}\label{sec:11.2}

 Similar to what we did in \citetalias{ber09}, we also calculate the distance
 to IC\,1613 using two methods based on the properties of the RR~Lyrae stars,
 namely the luminosity-metallicity relation, which arises from the knowledge
 that the intrinsic luminosity of HB stars mainly depends on their metallicity,
 and the period-luminosity-metallicity (PLM) relation, based on the theoretical
 location of the instability strip in the period-luminosity plane.

 The luminosity-metallicity relation we used in \citetalias{ber09} has the
 form:
\begin{displaymath}
 M_V = 0.866(\pm0.085) + 0.214(\pm0.047) [Fe/H].
\end{displaymath}
 To calculate the mean magnitude of the RR~Lyrae stars, we used only the stars
 for which we could determine accurate intensity-averaged mean
 magnitudes---RR$ab$ and RR$c$---and obtained
 $\langle V \rangle$=24.99$\pm$0.01.
 Assuming a mean metallicity of Z=0.0005$\pm$0.0002 (i.e.,
 [Fe/H]=$-$1.6$\pm$0.2 assuming Z$_\sun$=0.0198 \citep{gre93} and
 [Fe/H]=log Z+1.70 $-$ log(0.638 f + 0.362) \citep{sal93} with the
 $\alpha-$enhancement factor f set to zero) for the old population from the
 results of the SFH derived in Skillman et al.\ (2010, in preparation), we find
 a luminosity for the HB of M$_V$=0.52$\pm$0.12. The uncertainty was quantified
 through Monte-Carlo simulations. This gives a distance modulus of
 24.47$\pm$0.12, or (m$-$M)$_0$=24.39$\pm$0.12 after correcting for reddening.

 The second method we used to determine the distance modulus consists of
 matching the PLM relation at the first-overtone blue edge (FOBE) of the IS
 from \citet[their eq. 3]{cap00} to the observed RR$c$.
 Basically, the theoretical limits of the IS are shifted in magnitude until the
 FOBE coincides with the observed distribution of first-overtone RR~Lyrae
 stars. Figure~\ref{fig:15} shows the position of the IS ({\it dashed lines})
 overplotted on the distribution of observed RR~Lyrae stars in the $\langle V
 \rangle$--log~P plane.

 Adopting the mean metallicity given above (Z=0.0005), we derive a dereddened
 distance modulus of (m$-$M)$_0$=24.36, for which \citet{cap00} give an
 intrinsic dispersion of $\sigma _V$=0.07 mag due to uncertainties associated
 with the various ingredients of the model.
 Combined with the errors in metallicity, mean magnitude, and period, we
 estimate a total uncertainty in this measure of the distance of $\sim$0.1.

\subsection{Results}\label{sec:11.3}

 Since the distances derived in \S~\ref{sec:11.1} and \ref{sec:11.2} are
 consistent with each other and have similar uncertainties, we simply averaged
 them to obtain our best distance estimate and obtained a dereddened distance
 modulus of 24.44 ($\sigma$=0.054), corresponding to a distance of 770~kpc.

 A summary of the previous distance determination from the literature is given
 in Table~\ref{tab8}. The columns give the reddened distance modulus, the
 assumed or calculated reddening and LMC distance modulus, the observation
 band, and the method used to estimate the distance.
 We compare these values in Fig.~\ref{fig:16}, where the reddened moduli as
 given in the referenced articles are shown as pluses. To make this comparison
 more meaningful, we also show the distance moduli corrected for a common
 reddening of E(B$-$V)=0.025 \citep{sch98} and LMC distance modulus
 (m$-$M)$_0$=18.515$\pm$0.085 \citep{cle03}. The reddening correction was done
 following the extinction law of \citet{car89} with R$_V$=3.1.

 The figure shows that all the measurements, independent of the method, are in
 excellent agreement with each other once set on the same scale.
 From these values (excluding the \citet{sah92} outlier), we find a mean
 dereddened distance of (m$-$M)$_0$=24.400$\pm$0.014, where the uncertainty only
 includes the standard deviation of the different measurements.
 The vertical lines in Fig.~\ref{fig:16} indicate the mean and standard
 deviation ($\sigma$=0.071) of these values.
 Therefore, we believe that the largest source of uncertainty on the distance to
 IC\,1613 is systematic rather than statistical, and mostly lies in the
 calibration of the distance to the LMC.

\section{Discussion and Conclusions}\label{sec:12}

 We have presented the results of a new search for variable stars in IC\,1613
 based on high quality {\it HST} images.
 In the ACS field, we found 259 candidate variables, including 90 RR~Lyrae
 stars, 49 Cepheids, and 38 eclipsing binaries, as well as nine RR~Lyrae stars
 and two additional Cepheids in the parallel WFPC2 field. Only thirteen of
 these variables were known prior to this study, all of them Cepheids.

 We find that the mean periods of the RR$ab$ and RR$c$ stars, as well as the
 fraction of overtone pulsators, place this galaxy in the intermediate regime
 between the Oosterhoff types, as was already observed for the vast majority of
 Local Group dwarf galaxies. 

 From the comparison with Magellanic Clouds Cepheids through the
 period-luminosity diagram, we find that all our Cepheids are bona fide
 classical Cepheids, while the Cepheids in the dSph Cetus and Tucana discussed
 in Paper~I were classified as anomalous Cepheids. The lack of classical
 Cepheids in the latter dwarfs is not surprising given the absence of
 significant star formation more recent than about 8-10~Gyr \citep{mon10}. On
 the other hand, the reason for the apparent lack of anomalous Cepheids in
 IC\,1613 is not clear. The star formation histories calculated from deep CMDs
 \citep[2010, in preparation]{ski03} indicate that it formed stars fairly
 constantly over the past 13~Gyr, so unless the formation of anomalous Cepheids
 is extremely sensitive to the metal content, anomalous Cepheids should also be
 present in IC\,1613.

 \citet{gal04} showed that both types of Cepheids were present in
 comparable numbers in the transition type galaxy Phoenix, and possibly also in
 Leo A and Sextans A, all of them having very low metallicity (Z$\la$0.0008).
 These authors suggest that the only requirements for the presence of both
 types of Cepheids in a galaxy are low enough metallicity (Z$\sim$0.0004) and
 star formation activity at all ages.
 At the other end of the dwarf galaxy metallicity spectrum, 83 anomalous
 Cepheids were discovered in the LMC \citep{sos08b}, compared to the 3361
 classical Cepheids that are present in the same catalog \citep{sos08a}. This
 implies about 2.4 anomalous Cepheid per 100 classical Cepheid at the
 metallicity of the LMC (Z$\sim$0.008).
 If this fraction is representative for low metallicity galaxies, about one
 anomalous Cepheid could be expected in our ACS field-of-view of IC\,1613.
 However, \citet{dol02} showed how low metallicity galaxies contain many more
 short-period classical Cepheids than higher metallicity galaxies at a given
 star formation rate due to morphology of the blue loops: at low metallicity,
 the blue loops extend further to the blue and thus cross the instability strip
 at fainter magnitudes. As a result of the stellar initial mass function, the
 fainter Cepheids are also more numerous.
 Therefore, the apparent lack of anomalous Cepheids in our field might simply be
 due to small number statistics. Given the low density of anomalous Cepheids
 expected in IC\,1613, more data covering a large field-of-view are needed to
 check if anomalous Cepheids are actually present in this galaxy.

 Finally, we used the properties of the RR~Lyrae stars and Cepheids to estimate
 the distance to IC\,1613, and found excellent agreement with the values
 previously determined. Combining all the measurements after correction for a
 common reddening and reference LMC distance, we find a true distance modulus
 to this galaxy of (m$-$M)$_0$=24.400$\pm$0.014 (statistical), corresponding to
 760~kpc.

\acknowledgments

{\it Facility:} \facility{HST (ACS, WFPC2)}

 The authors would like to thank the anonymous referee for useful comments.
 Support for this work was provided by a Marie Curie Early Stage Research
 Training Fellowship of the European Community's Sixth Framework Programme
 (contract MEST-CT-2004-504604), the IAC (grant 310394), the Education and
 Science Ministry of Spain (grants AYA2004-06343 and AYA2007-3E3507), and NASA
 through grant GO-10505 from the Space Telescope Science Institute, which is
 operated by AURA, Inc., under NASA contract NAS5-26555.
 This research has made use of the NASA/IPAC Infrared Science Archive, which
 is operated by the Jet Propulsion Laboratory, California Institute of
 Technology, under contract with the National Aeronautics and Space
 Administration.

\appendix

\section{Comments on Individual Variables}

 The following comments are based on the careful inspection of the stacked
 images, of the light curves, and/or peculiar properties exhibited on one or
 more of the plots presented in this work. \\

  V001 --- Blend.

  V004 --- Blend.

  V007 --- Close to a bright star.

  V009 --- Blend.

  V010 --- Possible blend.

  V013 --- Close to the wing of a very bright star.

  V019 --- Possible blend.

  V020 --- Blend.

  V022 --- Blend.

  V027 --- Blend.

  V029 --- Blend.

  V030 --- Blend.

  V032 --- Blend.

  V034 --- Blend.

  V036 --- Blend and close to a bright star.

  V038 --- Blend. Very low amplitude RR$ab$.

  V043 --- Possible blend.

  V044 --- Possible blend.

  V046 --- Possible blend.

  V047 --- Blend.

  V050 --- Possible blend.

  V052 --- Possible blend.

  V057 --- Possible blend.

  V058 --- Blend.

  V061 --- Blend.

  V066 --- Blend.

  V071 --- Close to a very bright star.

  V072 --- Blend.

  V073 --- Blend.

  V074 --- Blend with a background galaxy and bright stars.

  V076 --- Blend.

  V077 --- Blend.

  V080 --- In bad column.

  V085 --- Blend.

  V087 --- Blend.

  V088 --- Blend.

  V089 --- Possible blend.

  V090 --- Blend.

  V093 --- Blend.

  V099 --- Located close to the edge of chip 1.
           Some points bad or missing because of dithering.

  V102 --- Blend.

  V106 --- Blend.

  V108 --- Blend.

  V114 --- Possible blend.

  V121 --- Blend and located close to the edge of chip 1.
           Some points bad or missing because of dithering.

  V122 --- Blend.

  V125 --- Blend.

  V133 --- Blend.

  V141 --- Blend.

  V144 --- Blend.

  V145 --- Possible blend.

  V148 --- Blend.

  V149 --- Blend.

  V152 --- Blend.

  V153 --- Blend.

  V161 --- Blend.

  V167 --- Blend.

  V170 --- Blend.

  V171 --- Blend.

  V175 --- Possible blend.

  V176 --- Blend.

  V177 --- Blend.

  V178 --- Blend.

  V180 --- Blend.

 VC202 --- Blend.

 VC206 --- Blend.

 VC208 --- Saturated in F814W.

 VC213 --- Possible blend.

 VC217 --- Blend.

 VC223 --- Blend.

 VC225 --- Possible blend.

 VC227 --- Blend.

 VC229 --- Blend.

 VC231 --- Blend.

 VC236 --- Blend.

 VC238 --- Blend.

 VC239 --- Blend.

 VC240 --- Blend.

 VC241 --- Blend.

 VC243 --- Blend.

 VC244 --- Possible blend.

 VC245 --- Possible blend.

 VC246 --- Possible blend.

 VC252 --- Blend.

 VC254 --- Blend.

 VC257 --- Possible blend.

 VC259 --- Blend.

 VC265 --- Blend.

 VC267 --- Blend.

 VC268 --- Blend.

 VC270 --- Blend.

\section{Finding Charts}

\begin{figure*}
\epsscale{1.0} 
\plotone{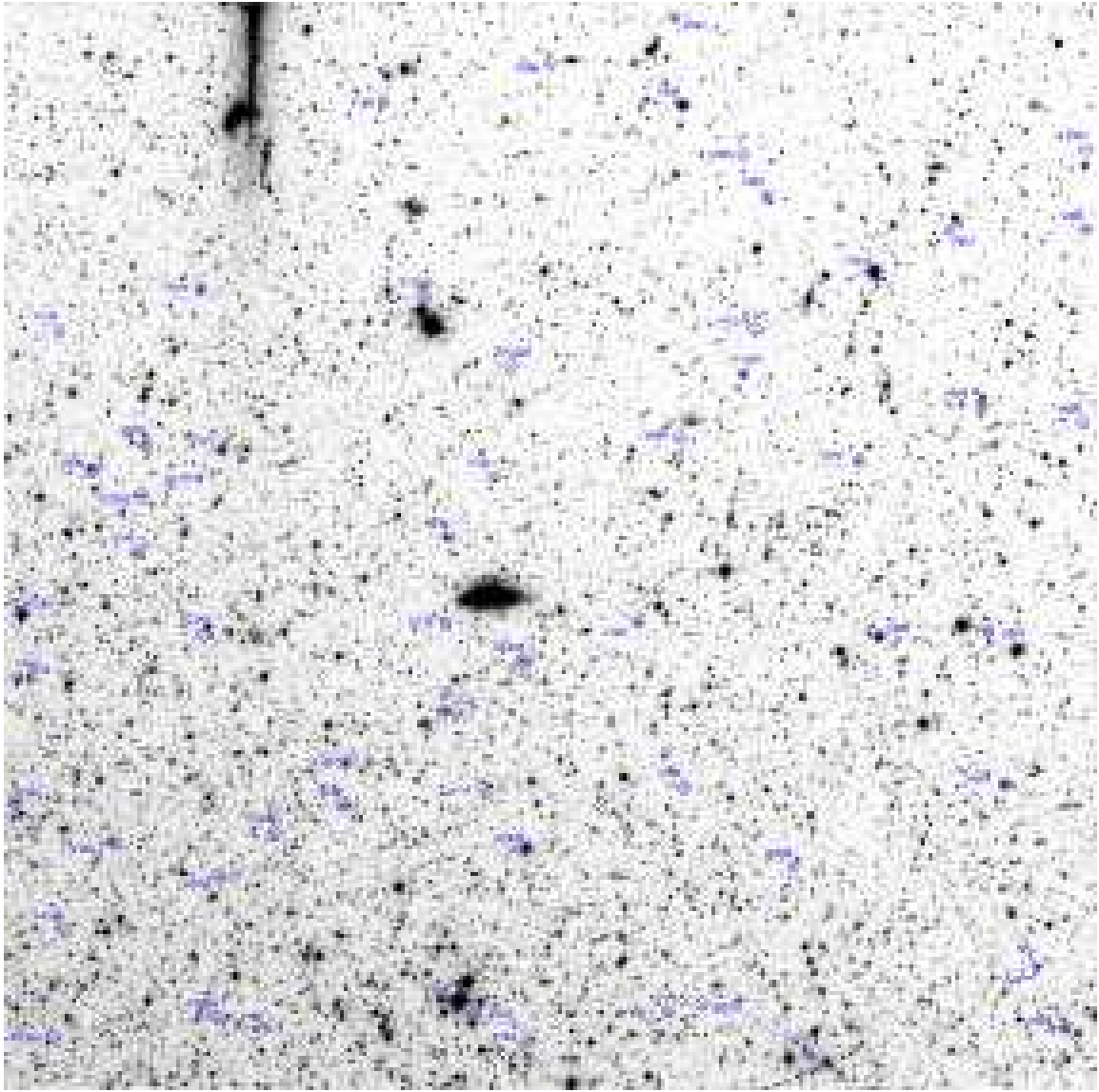}
\figcaption{Finding chart for the North-East quadrant in IC\,1613. The other
 quadrants are in the electronic edition. Variables and candidate variables are
 shown as open circles and open diamonds, respectively.
\label{fig:17}}
\end{figure*}

 The finding charts for the whole sample of variable stars are presented
 in the electronic version of The Astrophysical Journal (Fig.~\ref{fig:17}).

%
%

\end{document}